\documentclass[epsf]{aa}
\usepackage{graphicx}
\begin{document}
   \title{A new spectroscopic and interferometric study of the young stellar object V645\,Cyg\thanks{Partially
   based on data obtained at the 6--m BTA telescope of the Russian Academy of Sciences, 3.6--m Canada-France-Hawaii
   telescope, and 3--m Shane telescope of the Lick Observatory.}}

   \author{A.~S. Miroshnichenko\inst{1,2}
          \and
          K.-H. Hofmann\inst{1}, D. Schertl\inst{1}, G. Weigelt\inst{1}, S. Kraus\inst{1}
          \and
          N. Manset\inst{3}, L. Albert\inst{3}
          \and
          Y. Y. Balega\inst{4}, V. G. Klochkova\inst{4}
          \and
          R. J. Rudy\inst{5}, D. K. Lynch\inst{5}, S. Mazuk\inst{5}, C. C.
          Venturini\inst{5}, R. W. Russell\inst{5}
          \and
          K. N. Grankin\inst{6}
          \and
          R. C. Puetter\inst{7}
          \and
          R. B. Perry\inst{8}
          }

   \offprints{A.~S. Miroshnichenko}

   \institute{Max-Planck-Institut f\"ur Radioastronomie, Auf dem H\"ugel 69, 53121 Bonn, Germany\\
         \and
             University of North Carolina at Greensboro, Department of Physics and Astronomy, P.O. Box
             26170, Greensboro, NC 27402--6170, USA,
             \email{a$\_$mirosh@uncg.edu}\\
         \and
             CFHT Corporation, 65--1238 Mamalahoa Hwy, Kamuela, HI 96743, Hawaii, USA\\
         \and
             Special Astrophysical Observatory of the Russian Academy of Sciences, Nizhnyj Arkhyz, 369167, Russia\\
         \and
             The Aerospace Corp. M2/266, P.O.Box 92957, Los Angeles, CA 90009, USA\\
         \and
             Crimean Astrophysical Observatory, Nauchny, Crimea 334413, Ukraine\\
         \and
             Center for Astrophysics and Space Science, University of California, San Diego, C--0111, La Jolla,
             CA 92093, USA\\
         \and
             Earth and Space Science Support Office, M/S 160, NASA Langley Research Center, Hampton, VA
             23681, USA
             }

   \date{Received date; accepted date}


  \abstract
   {}
   {We present the results of high-resolution optical spectroscopy, low-resolution near-IR spectroscopy
   and near-infrared speckle interferometry of the massive young stellar object candidate V645\,Cyg, acquired
   to refine its fundamental parameters and the properties of its circumstellar envelope.}
   {Speckle interferometry in the $H$- and $K$-bands and an optical spectrum in the range 5200--6680 \AA\ with a
   spectral resolving power of $R$ = 60 000 were obtained at the 6\,m telescope of the Russian
   Academy of Sciences. Another optical spectrum in the range 4300--10500 \AA\ with $R$ = 79 000 was
   obtained at the 3.6\,m CFHT. Low-resolution spectra in the ranges 0.46--1.4 $\mu$m and 1.4--2.5 $\mu$m with
   $R \sim$ 800 and $\sim$ 700, respectively, were obtained at the 3\,m Shane telescope of the Lick Observatory.}
   {Using a novel kinematical method based on the non-linear modeling of the neutral hydrogen density profile in the direction
   toward the object, we propose a distance of $D = 4.2\pm$0.2 kpc. We also suggest a revised estimate of the
   star's effective temperature, T$_{\rm eff} \sim$25 000 K. We resolved the object in both $H$- and $K$-bands.
   Using a two-component ring fit, we derived a compact component size of 14 mas and 12 mas in the
   $H$- and $K$-band, respectively, which correspond to 29 and 26 AU at the revised distance.
   Analysis of our own and previously published data indicates a $\sim$2 mag decrease in the near-infrared brightness of
   V645\,Cyg at the beginning of the 1980's. At the same time, the cometary nebular condensation N1
   appears to fade in this wavelength range with respect to the N0 object, representing the star with a nearly pole-on
   optically-thick disk and an optically-thin envelope.}
{We conclude that V645\,Cyg is a young, massive, main-sequence star,
which recently emerged from its cocoon and has already experienced
its protostellar accretion stage. The presence of accretion is not
necessary to account for the high observed luminosity of
(2--6)$\times$10$^4$ M$_{\odot}$\,yr$^{-1}$. The receding part of a
strong, mostly uniform outflow with a terminal velocity of $\sim$800
km\,s$^{-1}$ is only blocked from view far from the star, where
forbidden lines form. The near-infrared size of the source is
consistent with the dust sublimation distance close to this hot and
luminous star and is the largest among all young stellar objects
observed interferometrically to-date.}

   \keywords{techniques: spectroscopic, interferometric -- stars: early-type, circumstellar matter
   -- stars: winds, outflows -- stars: individual (V645\,Cyg)
               }

   \authorrunning{Miroshnichenko et al.}
   \titlerunning{Properties of V645\,Cyg}

   \maketitle
%

\section{Introduction}\label{intro}

The object was discovered as a variable star with long-term
wave-like variations by Hoffmeister, Rohlfs, \& Ahnert
(\cite{hra51}) and received its name of \object{V645\,Cyg} in the
General Catalog of Variable Stars (Kholopov et al. \cite{kh90}). A
strong IR source was detected at the same location in the course of
the CRL survey (\object{CRL\,2789}, Walker \& Price \cite{wp75}) and
identified with an optically faint, starlike nebula (N0),
accompanied by a cometary shape structure (N1) a few arcseconds
apart (Cohen \cite{c77}). Goodrich (\cite{g86}) discovered a more
complex structure in the object's optical image and called it the
``Duck nebula''.

Several initial studies of the object gave controversial results,
some of which are summarized in Table \ref{t1}. Absorption lines in
the spectrum of N1 were attributed to an expanding shell (Humphreys
et al. \cite{hmb80}). No evidence of the nature of the underlying
star was found. The reddening derived from the optical color-indices
was attributed to the interstellar (IS) medium.

\begin{table}
\caption{Published results for the nature and evolutionary state of
V645\,Cyg} \label{t1} \centering
\begin{tabular}{ccclll}
\hline\noalign{\smallskip}
Sp.   & A$_V$ & $V$ & $D$  & Ref.         \\
type  & mag   & mag & kpc  &              \\
\noalign{\smallskip}\hline\noalign{\smallskip}
O7    & 4.2   & 15.1$^{\rm a}$&$\ge$3,$\sim$6& Cohen (\cite{c77})\\
      & 3.6   &     &$\ge$1, $\le$6& Humphreys et al. (\cite{hmb80})\\
A0    & 2.5   &     &3.5$\pm$0.5   & Goodrich (\cite{g86})\\
\noalign{\smallskip}\hline
\smallskip
\end{tabular}
\begin{list}{}
\item Estimate of the star's spectral type is listed in column
1; interstellar extinction toward the object in column 2; object's
visual brightness in column 3; distance toward the object in column
4; and a reference to the estimates in column 5
\item $^{\rm a}$ the brightness for the knot N0
\end{list}
\end{table}

Hamann \& Persson (\cite{hp89}) used medium-resolution optical
spectroscopy and concluded that N0 and N1 represent reflected light.
Analyzing line profiles in their spectra, they proposed the presence
of a circumstellar (CS) disk and a bipolar outflow. Distinctive
features of the spectrum are blue-shifted, forbidden, emission lines
of [O {\sc I}] and [S {\sc II}], which can be explained by blocking
the receding part of the outflow by the optically-thick disk. Later
these authors suggested that V645\,Cyg might be a FU\,Ori type star
with an accretionally heated disk (Hamann \& Persson \cite{hp92}).
The object's morphology (a central condensation accompanied by an
arc-shaped filament), sometimes referred to as cometary nebula, is
sometimes observed in FU\,Ori type objects (e.g., \object{Z\,CMa})
as well as in Herbig Be candidate stars (e.g., \object{MWC\,137}).
Thus, it cannot by itself be an indicator of the object's type.

Photometric data (see Sect. \ref{discussion}) imply that V645\,Cyg
has had no FU\,Ori type outburst in the past $\sim$60 years.
Moreover, optical color-indices of V645\,Cyg are typical for
reddened early B-type stars, while FU\,Ori type objects have
color-indices typical of F-type supergiants. On the other hand,
blue-shifted forbidden lines, similar to those of V645\,Cyg, are
seen in the spectrum of Z\,CMa (a FU\,Ori candidate associated with
a similar cometary nebula), which shows bluer colors at maximum
brightness. The latter are due to a variable contribution of a
high-excitation source (early B-type) in addition to a mid F-type
supergiant spectrum.

V645\,Cyg is a prominent IR source detected by the IRAS, ISO, and
MSX satellites. The mid-IR spectra exhibit a $\lambda$10$\mu$m
silicate feature in absorption (shown in Fig. \ref{f9} below), which
may be attributed to the IS extinction (Bowey, Adamson, \& Yates
\cite{bay03}). Lorenzetti et al. (\cite{ltg99}) reported detection
of forbidden emission lines of [O {\sc I}] at 63.2 $\mu$m, [O {\sc
III}] at 88.5 $\mu$m, and of [C {\sc II}] at 157.8 $\mu$m in the ISO
spectrum of the object. They found that these lines are also present
in the spectra of Herbig Ae/Be stars and that the [O {\sc III}] line
was associated only with the hottest objects (earlier than B0).

Near-IR imaging polarimetry (Minchin et al. \cite{mhby91}) shows
that the polarization decreases with wavelength from $\sim$10\% in
the $J$-band to $\sim$3.6\% in the $K$-band. The polarization
position angle rotates significantly with wavelength (9$^{\circ}$ in
the $J$-band to 36$^{\circ}$ in the $K$-band). These results were
interpreted as evidence of a disk around N0 and N1 being a slab of
material swept-up by the outflow from N0.

Variable methanol (Slysh et al. \cite{svk99}, Szymchak, Hrynek, \&
Kus \cite{shk00}, Blaszkiewicz \& Kus \cite{bk04}) and water vapour
(Comoretto et al. \cite{cpc90}) maser emission from the source was
detected in several studies. These types of masers are common in
massive young stars.

Clarke et al. (\cite{cl06}, hereafter C06) published low-resolution
optical and near-IR spectra, near/mid IR imaging, and $^{13}$CO line
observations of V645\,Cyg. They found redshifted (1800--2000 km\,
s$^{-1}$), weak, emission features close to the H$\alpha$ and He
{\sc I} $\lambda$1.701 $\mu$m lines, and the CO $\lambda$2.29 $\mu$m
band. Their presence was interpreted as a possible manifestation of
the receding part of the outflow. They suggested that the star is
surrounded by a pseudo-photosphere, which implies that the object
resembles a late O-type supergiant obscuring a hotter star inside.
They also concluded that N0 coincides with the position of the star.
No emission from the knot N1 was observed at $\lambda 10 \mu$m.

Despite a general consensus that V645\,Cyg is a young stellar
object, there is still no agreement about its physical parameters.
Some authors prefer the O-type classification (e.g., Testi, Palla,
\& Natta \cite{tpn98}), while others adopt the A-supergiant idea
(e.g., Bowey et al. \cite{bay03}). The distance toward the object is
also not well-constrained.

In this paper, we attempt a revised analysis of the properties of
this remarkable object based on high-resolution ($R \ge$ 60000)
optical spectroscopy and near-IR speckle interferometry, which have
been obtained for the first time.

Our observations and results are described in Sects. \ref{observ}
and \ref{results}, respectively. Analysis of our and the other
available data is presented in Sect. \ref{discussion}, and our
conclusions, predictions, and suggestions for future observations
are given in Sect. \ref{conclus}.

\section{Observations}\label{observ}

The speckle interferograms of V645\,Cyg were recorded in September
2002 with the 6\,m telescope of the Special Astrophysical
Observatory (SAO) of the Russian Academy of Sciences (see Table
\ref{t2}). The detector of our speckle camera was a Rockwell HAWAII
array. A $K$-band filter with a central wavelength of
$\lambda$2115\,nm and a bandwidth 214\,nm, and a $H$-band filter
with a central wavelength of $\lambda$1648\,nm and a bandwidth
317\,nm were used. In the $K$- and $H$-band observations, the sizes
of one pixel on the sky corresponded to 27\,mas and 20\,mas,
respectively. Further observational parameters are listed in Table
\ref{t2}.

\begin{figure*}[ht]
\centering
\begin{tabular}{cc}
\hspace*{0.5cm}\resizebox{5.20cm}{!}{\includegraphics{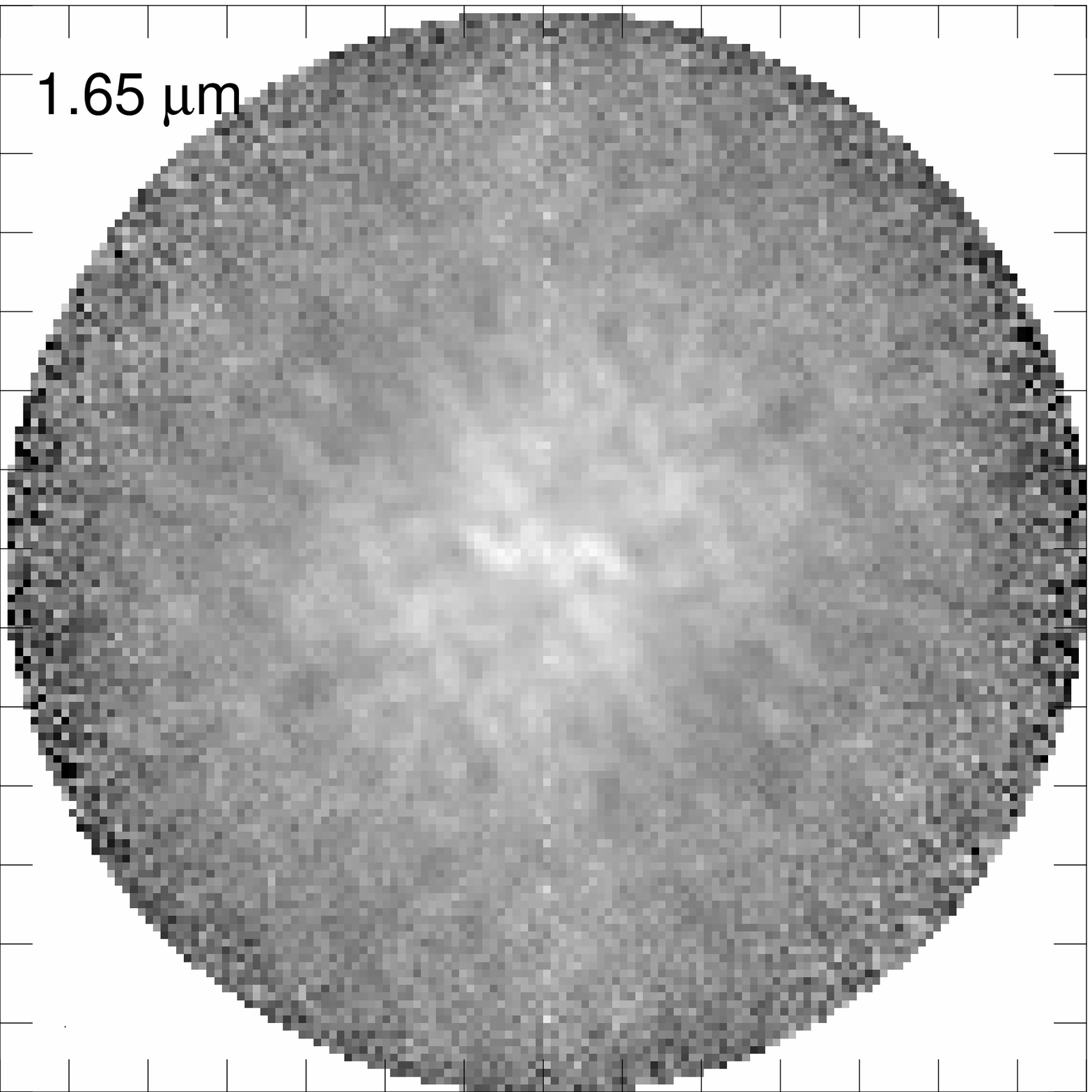}} &\hspace*{0.3cm} \resizebox{5.25cm}{!}{\includegraphics{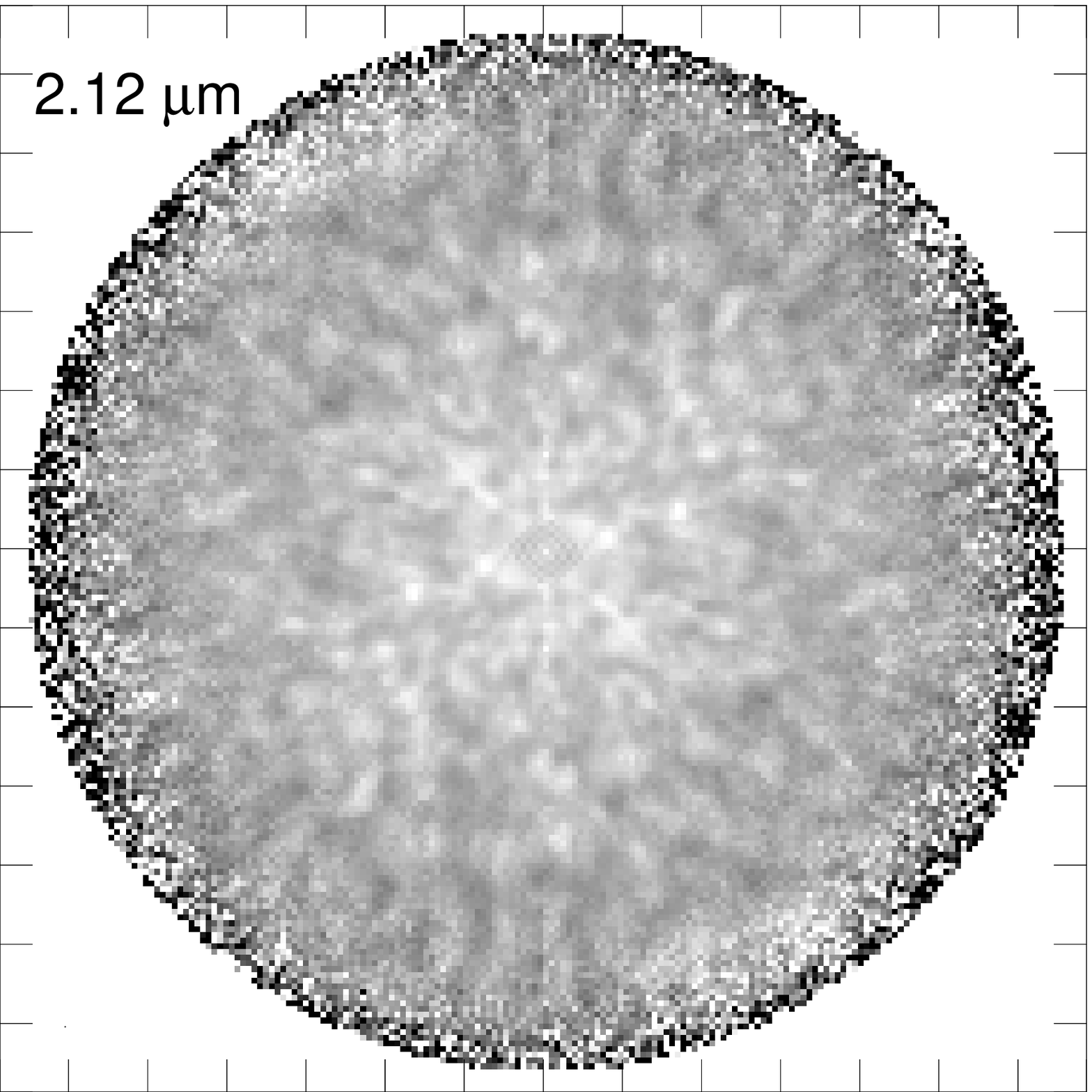}}\\
&\\
&\\
$\!\!\!$\resizebox{6.80cm}{!}{\includegraphics{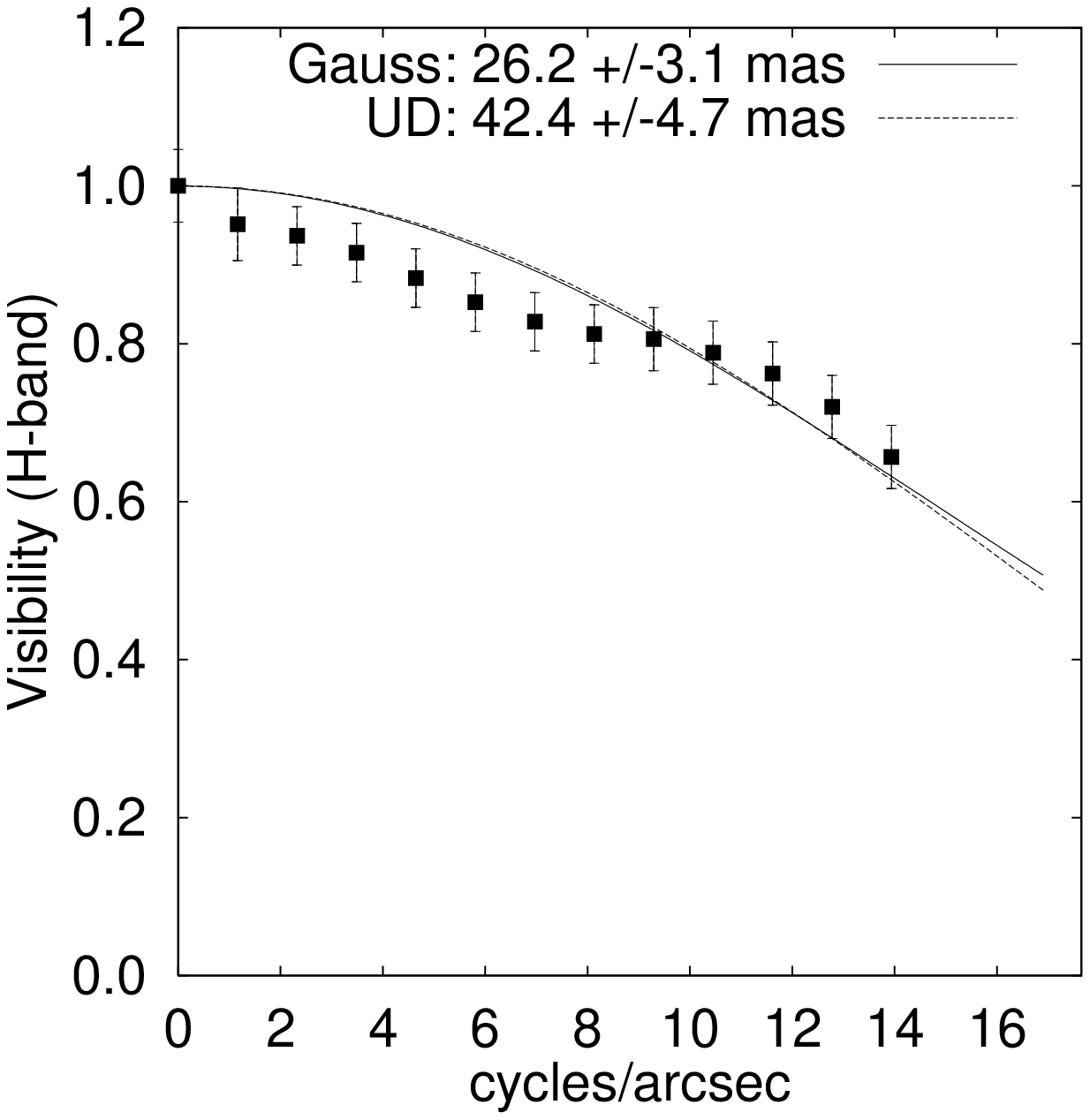}}& $\!\!\!$\resizebox{7.3cm}{!}{\includegraphics{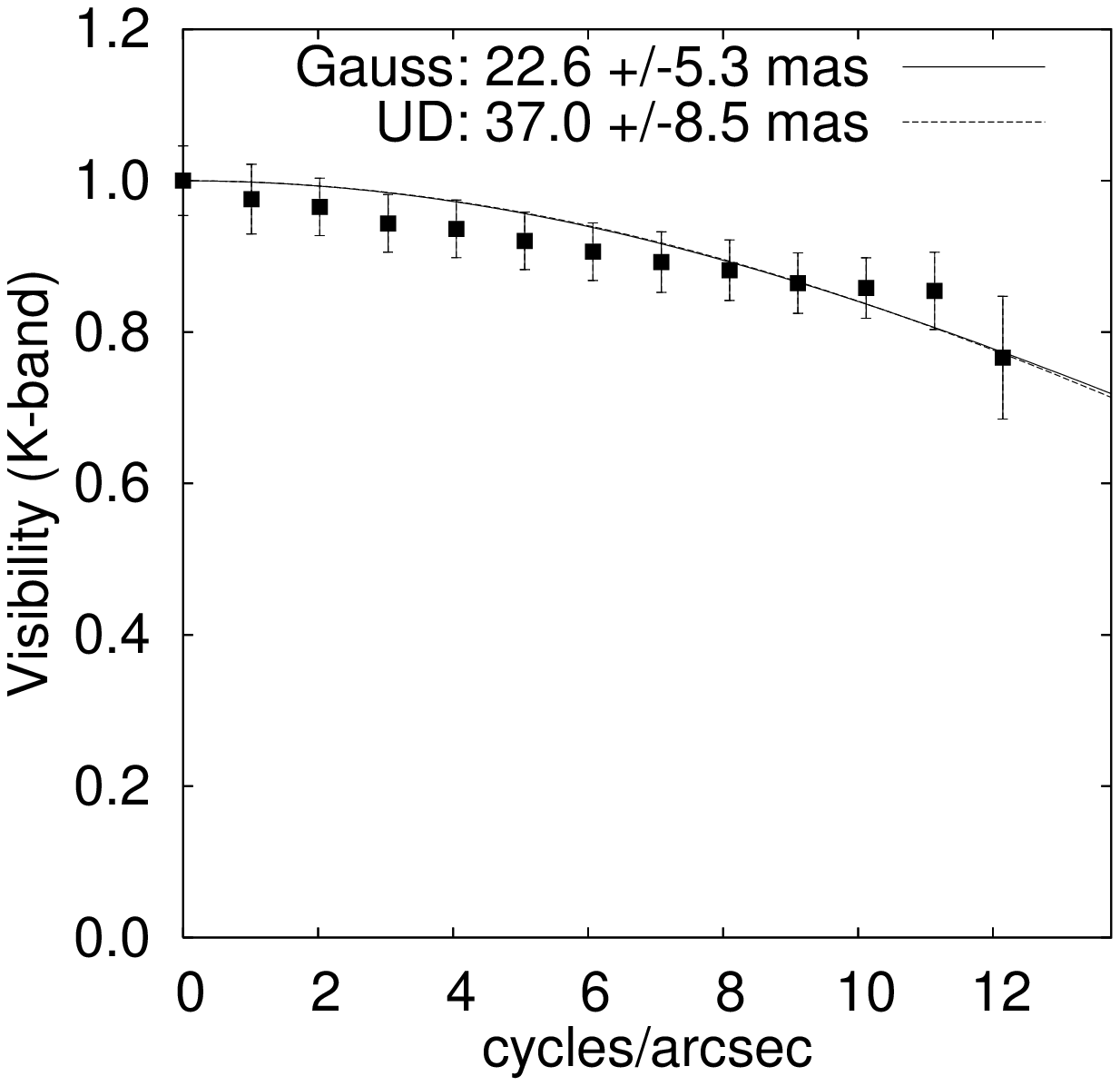}}\\
$\!\!\!$\resizebox{6.8cm}{!}{\includegraphics{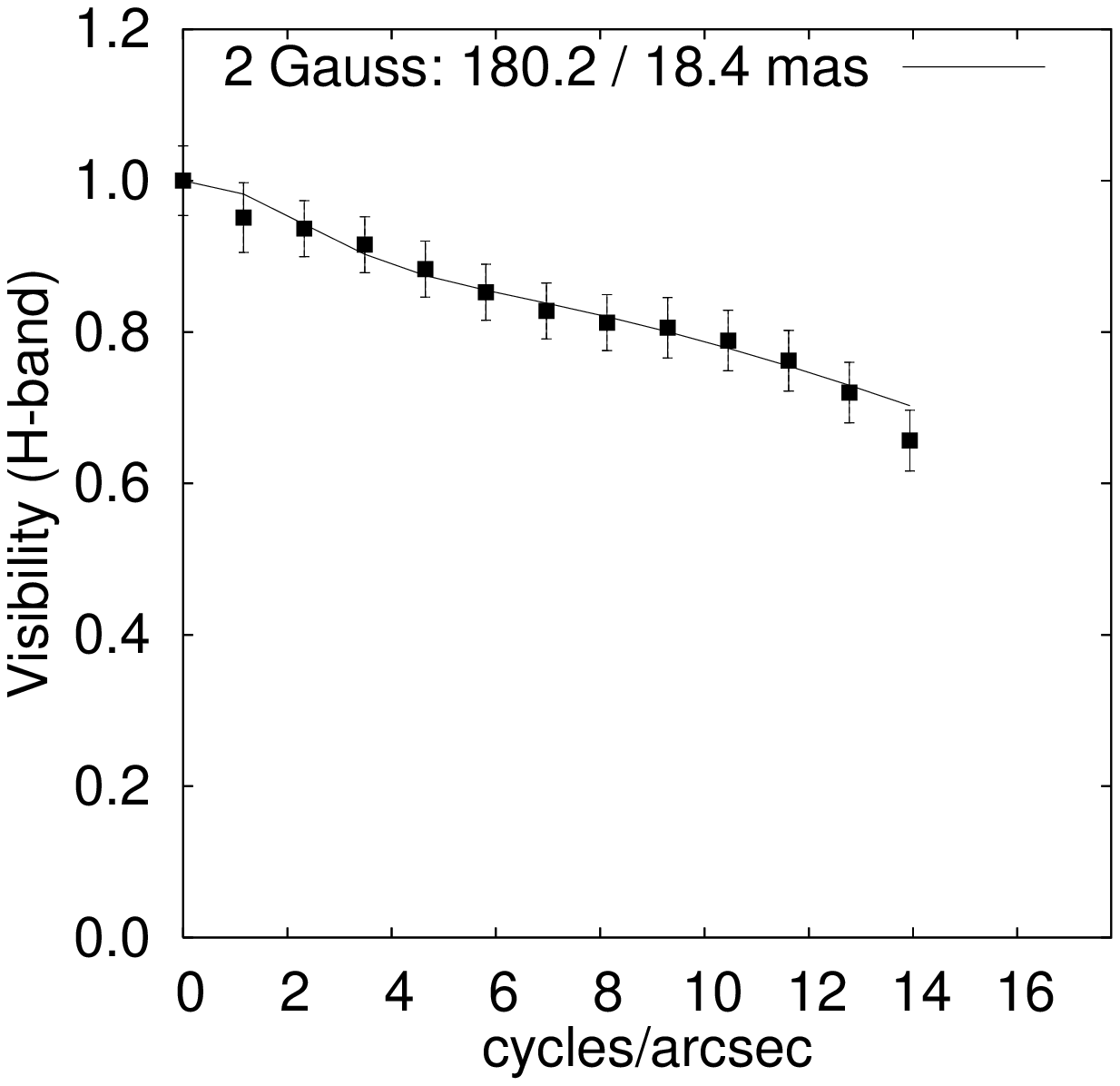}} & $\!\!\!$\resizebox{6.8cm}{!}{\includegraphics{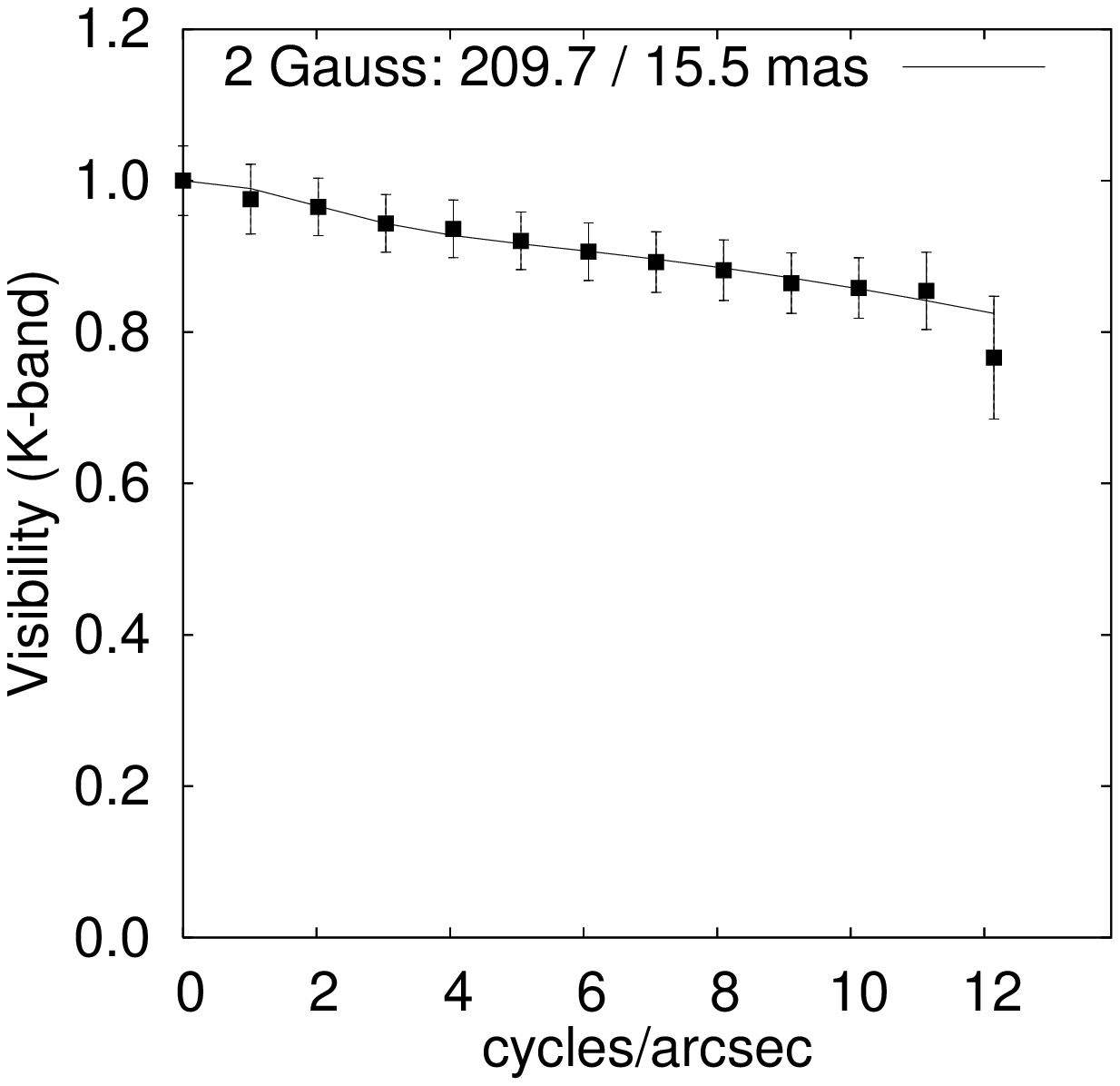}}\\
\end{tabular}
\caption{Measured diffraction-limited two-dimensional visibilities
of V645\,Cyg in the $H$-band (top left) and $K$-band (top right) as a
function of spatial frequency in cycles/arcsec on both axes. Radial
dependence of the azimuthally averaged two-dimensional visibilities
is shown together with one-component Gaussian and uniform disc (UD)
fits (middle panels) as well as two-component Gaussian fits (lower
panels). The $H$-band data are presented in the left column, and the
$K$-band data in the right column. The fit range is up to the
telescope cut-off frequency. The numbers are the FWHM diameters of
the large/small component of the two-component model (see Table
\ref{t3}). } \label{f1}
\end{figure*}

\begin{figure*}[ht]
\centering
\begin{tabular}{cc}
\resizebox{7.20cm}{!}{\includegraphics{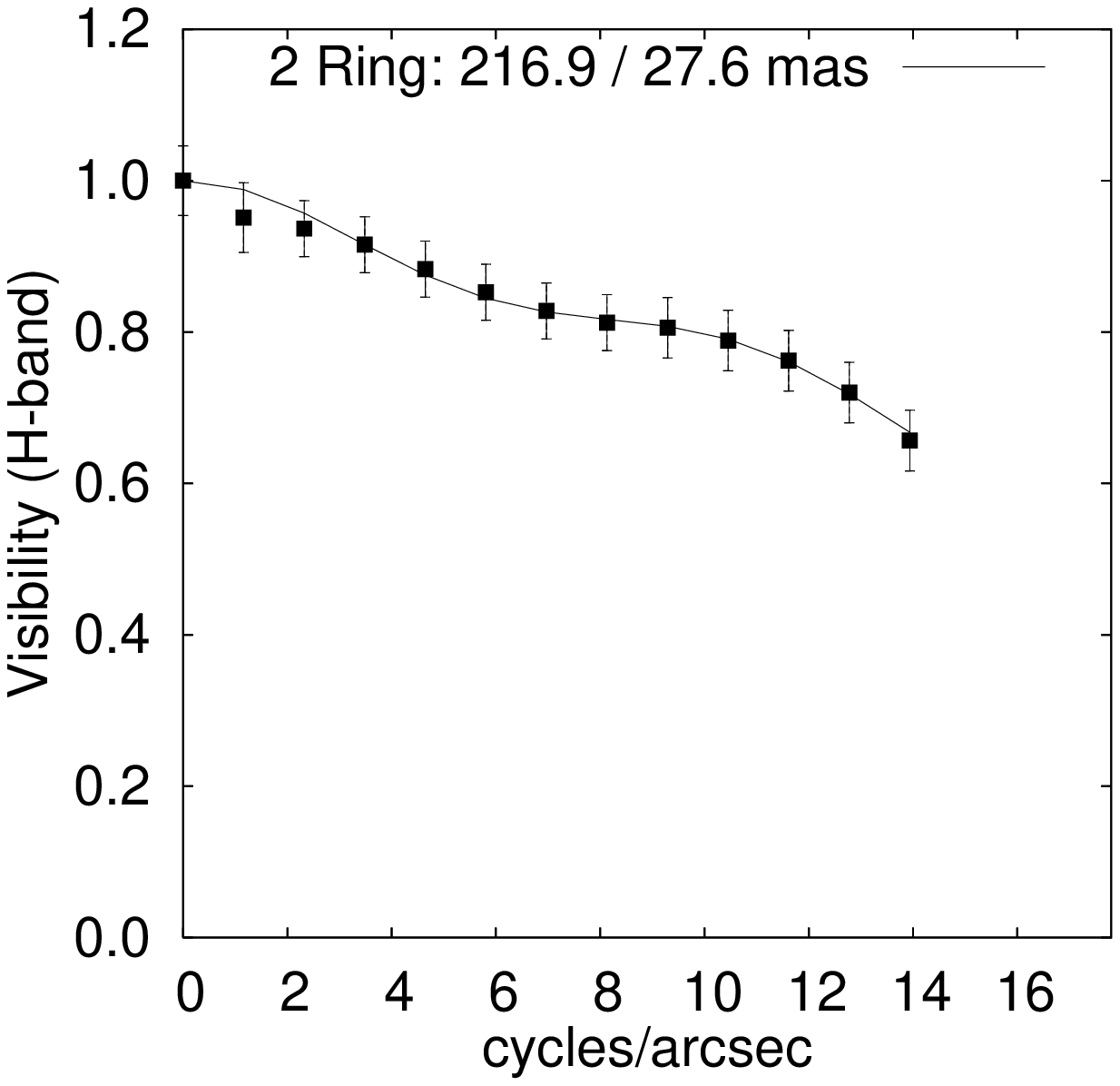}} & \resizebox{7.25cm}{!}{\includegraphics{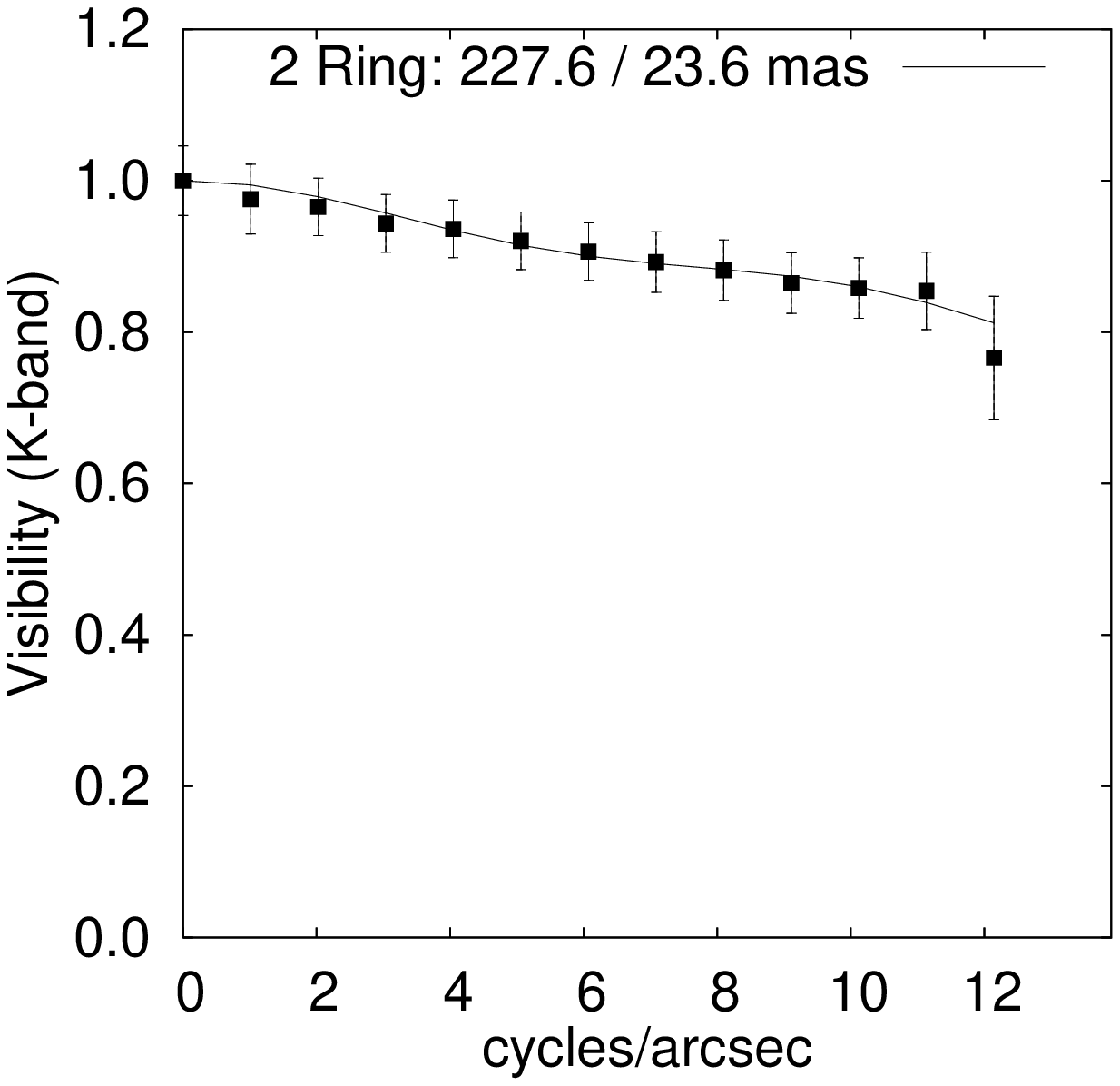}}\\
\end{tabular}
\caption{Visibilities of the best-fitting two-component models
consisting of two centered rings fitted to the azimuthally averaged
measured visibilities. The thickness of each ring was assumed to be
20\% of the ring diameter. The fit range is up to the telescope
cut-off frequency.
The listed numbers are the diameters of the large/small component of
the two-component model (see Table~\ref{t4}). } \label{f2}
\end{figure*}

The modulus of the Fourier transform of the object (visibility) was
obtained with the speckle interferometry method (Labeyrie
\cite{lab70}). The speckle interferograms of the unresolved stars
HD\,206824 (in the $K$-band) and BD+48$\degr$3475 (in the $H$-band)
were recorded immediately before and after the object's
observations, and served to determine the speckle transfer function.
The reconstructed two-dimensional visibility of V645\,Cyg in both
filters and their model fits are shown in Figs. \ref{f1} and
\ref{f2}.
%
\begin{table}[ht]
\caption{Details of the speckle interferometry observations.}
\label{t2} \centering
\begin{tabular}{ccccccc}
\hline\noalign{\smallskip}
Date      &  Filter   &  Frame size    & $N_{\rm T}$  &  $N_{\rm R}$  & $t_{\rm exp}$  & Seeing     \\
2002      &           &    [pixel]     &              &               &  [ms]          & [$\arcsec$]\\
\noalign{\smallskip}\hline\noalign{\smallskip}
Sep. 23   &  $K$      & 300$\times$300 &  1200        & 1239          & 218            & 2.2        \\
Sep. 24   &  $H$      & 256$\times$256 &  1050        & 1400          & 180            & 1.5        \\
\noalign{\smallskip}\hline\noalign{\smallskip}
\end{tabular}
\begin{list}{}
\item $N_{\rm T}$ and $N_{\rm R}$ are the numbers of recorded speckle
interferograms of V645\,Cyg and the reference stars, respectively.
$t_{\rm exp}$ is the exposure time per frame.
\end{list}
\end{table}

\begin{table}[ht]
\caption{Two-component fit diameters (Gaussian functions). }
\label{t3} \centering
\begin{tabular}{cccc}
\hline\noalign{\smallskip}
 Filter      &  FWHM [mas]      & FWHM [mas]      &  Brightness Ratio      \\
             &  Large Comp. & Small Comp. &  Large/Small\\
\noalign{\smallskip}\hline\noalign{\smallskip}
$K$          &  210$\pm$30    & 15.5$\pm$3      &  0.069$\pm$0.012       \\
$H$          &  180$\pm$22    & 18.4$\pm$3      &  0.126$\pm$0.022       \\
\noalign{\smallskip}\hline
\end{tabular}
\end{table}

\begin{table}[ht]
\caption{Two-component fit diameters (ring functions).} \label{t4}
\centering
\begin{tabular}{cccc}
\hline\noalign{\smallskip}
 Filter      &  FWHM [mas]      & FWHM [mas]      &  Brightness Ratio      \\
             &  Large Comp. & Small Comp. &  Large/Small\\
\noalign{\smallskip}\hline\noalign{\smallskip}
$K$          &  228$\pm$20    & 23.6$\pm$4      &  0.046$\pm$0.015       \\
$H$          &  217$\pm$15    & 27.6$\pm$4      &  0.085$\pm$0.021       \\
\noalign{\smallskip}\hline
\end{tabular}
\end{table}

High-resolution optical spectroscopy of V645\,Cyg was obtained on
July 10, 2004 at the SAO 6\,m telescope, and in August 2006 at the
3.6\,m CFHT. At SAO, we used the \'echelle spectrometer NES (Panchuk
et al. \cite{pan07}), which provides a spectral resolving power of
$R$ = 60 000. The spectral range observed was 5280--6680 \AA. The
resulting spectrum was a sum of two one-hour exposures with an
average signal-to-noise (S/N) ratio of $\sim$30 per pixel. The
spectroscopic data were processed using standard procedures of the
MIDAS package.

\begin{figure*}[ht]
\centering
\begin{tabular}{cc}
\resizebox{9.0cm}{!}{\includegraphics{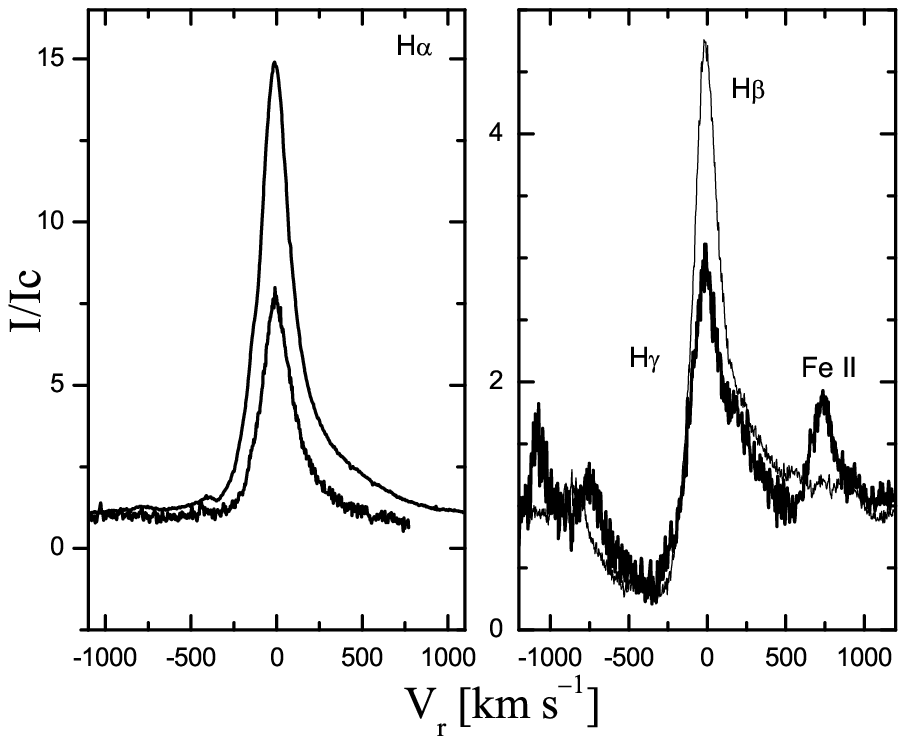}} & \resizebox{9.5cm}{!}{\includegraphics{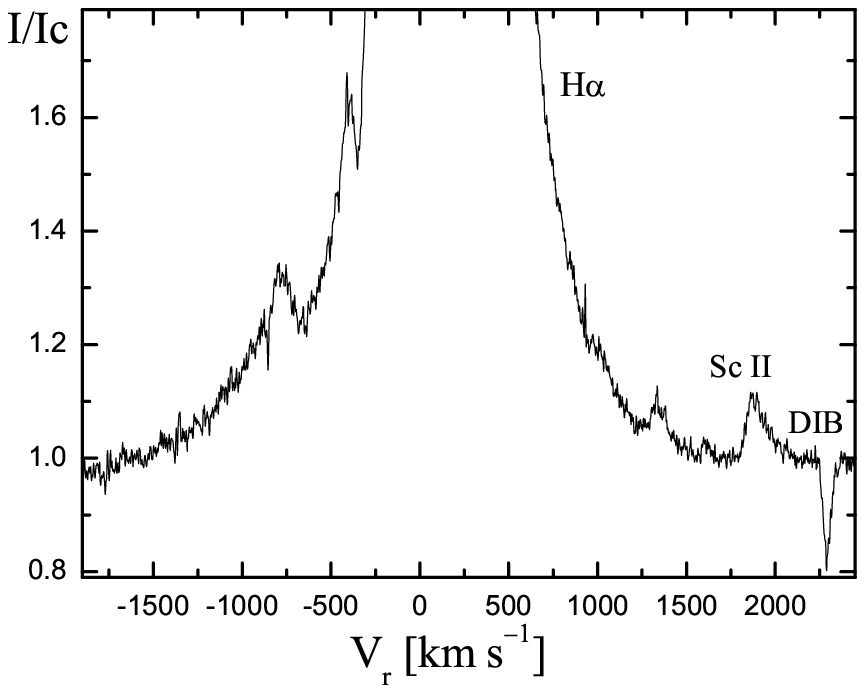}} \\
\end{tabular}
\caption{Portions of our CFHT and SAO spectra of V645\,Cyg. All the
portions are from the CFHT spectrum, except for the weaker H$\alpha$
line profile in the left panel. Unmarked features within the
H$\alpha$ line in the right panel may be due to clumps in the
outflow. The intensity is normalized to nearby continuum. The
velocity and wavelength scales are heliocentric.} \label{f3}
\end{figure*}

The spectroscopic mode of the \'echelle spectropolarimeter ESPaDOnS
(Manset \& Donati \cite{md03}) was used at CFHT. The resulting
spectrum was a sum of twelve individual spectra (five taken on 2006
August 6 and seven on 2006 August 8) of a total exposure time of
7970 seconds obtained with an average $R \sim$79 000 (per resolution
element, 2 pixels) in the range $\lambda\lambda$3600--10 500 \AA.
The S/N ratio gradually increased from $\sim$15 at H$\gamma$ to
$\sim$100 at H$\alpha$ due to the object's reddened SED. The data
reduction was completed with Libre-ESpRIT, written by J.-F. Donati
of Observatoire Midi-Pyrenees, and provided by CFHT. We identified
nearly 150 spectral lines in the range $\lambda\lambda$4320--8750
\AA, and present their parameters in Table \ref{t5}. The hydrogen
Paschen series lines were detected up to P22, but those that are
heavily contaminated by telluric absorption bands are not included
in the Table.

The low-resolution spectrum of V645\,Cyg was obtained at the 3\,m
Shane telescope of the Lick Observatory with the Near-InfraRed
Imaging Spectrograph (NIRIS, Rudy, Puetter, \& Mazuk \cite{rpm99})
on 2006 June 14. NIRIS has three channels and provides an almost
continuous coverage between 0.48 and 2.5 $\mu$m as well as a nearly
uniform dispersion within each channel. The average $R$ was
$\sim$800 in the optical channel (0.48 $\mu$m $\le \lambda \le 0.9
\mu$m) and $\sim$700 in both IR channels (0.8 $\mu$m $\le \lambda
\le 1.4 \mu$m and 1.4 $\mu$m $\le \lambda \le 2.5 \mu$m).

\section{Results}\label{results}

\subsection{Optical spectroscopy}\label{optical}

Qualitatively our high-resolution optical spectra are similar to
those published by Hamann \& Persson (\cite{hp89}) obtained in the
1980's and by C06 obtained in the 1990's. Even the emission line
widths (e.g., the Ca {\sc II} IR triplet and Fe {\sc I} lines, see
Fig. \ref{f3}) are virtually identical, providing evidence of the
long-term ($\sim$25 years) stability of the underlying star and its
CS structures.

\begin{figure}[ht]
\centering \resizebox{10cm}{!}{\includegraphics{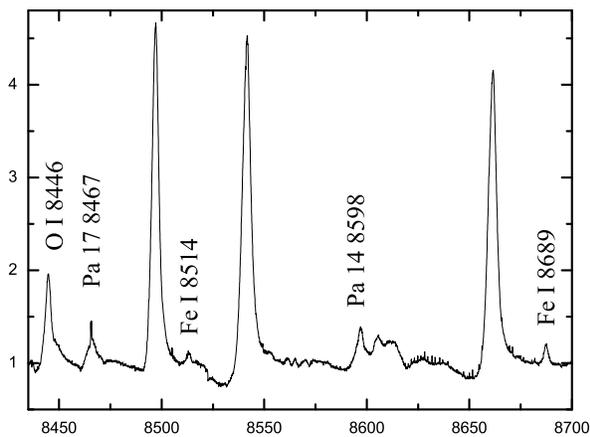}}
\caption{Red part of the CFHT spectrum of V645\,Cyg. The three
strong unmarked lines are blends of the IR Ca {\sc I} triplet and
hydrogen lines of the Paschen series. The intensity is normalized to
nearby continuum, and the wavelengths are heliocentric.} \label{f4}
\end{figure}

\begin{figure}[ht]
\centering \resizebox{9.8cm}{!}{\includegraphics{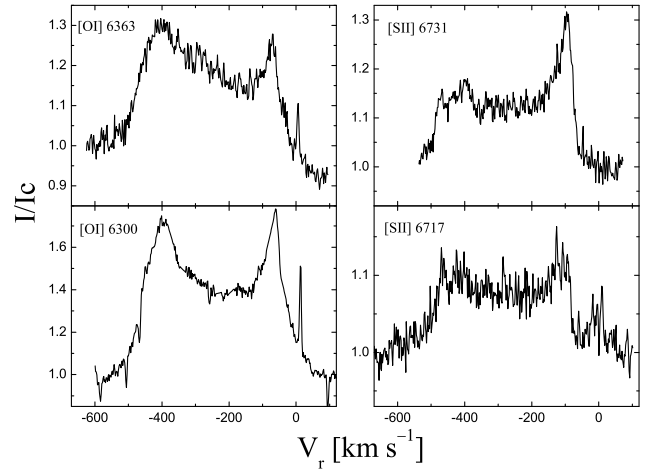}}
\caption{Blueshifted forbidden lines in the CFHT spectrum of
V645\,Cyg. The sharp emission components on the red wings of the [O
{\sc I}] lines are telluric. The intensity is normalized to nearby
continuum. The velocity scales is heliocentric. Similar profiles
were presented by Hamann \& Persson (\cite{hp89}) for the [S {\sc
II}] lines and by Acke, van den Ancker, \& Dullemond (\cite{aad05})
for the [O {\sc I}] lines.} \label{f5}
\end{figure}

\begin{figure*}[ht]
\centering \resizebox{14cm}{!}{\includegraphics{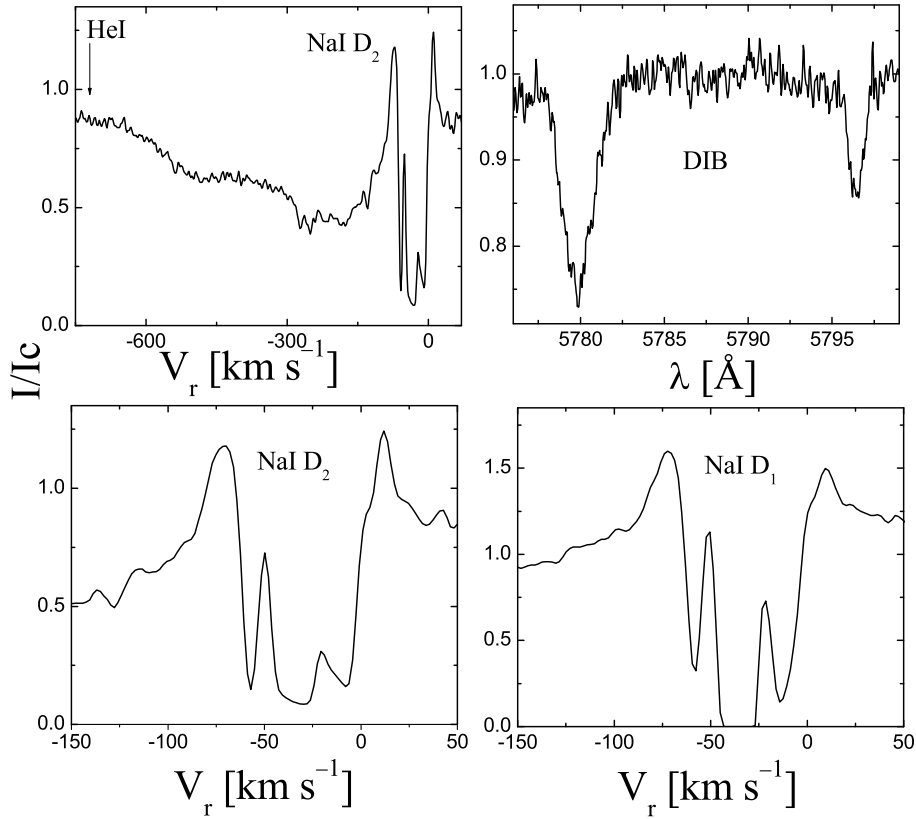}}
\caption{DIBs and sodium lines in the CFHT spectrum of V645\,Cyg.
Position of the He {\sc I} $\lambda$5876 \AA\ line, which is blended
with a wide and shallow CS component of the Na {\sc I} D$_2$ line,
is shown by the arrow. The lower panels show three IS components of
the Na {\sc I} D-lines. The intensity is normalized to nearby
continuum. The velocity and wavelength scales are heliocentric.}
\label{f6}
\end{figure*}

The combination of our spectral resolution (5--10 km\,s$^{-1}$) and
spectral coverage significantly exceeds in both quality and quantity
all previously published data for V645\,Cyg. This allowed us to
obtain for the first time a detailed profile of the H$\gamma$ line,
detect a faint He {\sc I} $\lambda$7065 \AA\ line (both have P Cyg
type profiles), and measure parameters of several diffuse IS bands
(DIBs). Portions of our CFHT and SAO optical spectra of V645\,Cyg
are shown in Figs. \ref{f3}--\ref{f6}.

The equivalent widths of the DIBs at $\lambda$5780 \AA\ and
$\lambda$5797 \AA\ (0.63 \AA\ and 0.20 \AA, respectively) are
consistent with a color-excess of E$(B-V)$ = 1.28$\pm$0.05 mag
(Herbig \cite{h93}). This result is in good agreement with E$(B-V)
\sim$ 1.35$\pm$0.10 mag, deduced from optical photometry (see Sect.
\ref{var}) by assuming an O- to early B-type for the underlying
star. Along with the conclusion of C06 that the position of N0
corresponds to the star's position, it also indicates a small amount
of CS reddening. Nevertheless, despite detecting weak DIBs (e.g.,
$\lambda$6699 \AA\ with an equivalent width of 0.02 \AA\ and an
intensity of 0.98 with respect to the continuum), no photospheric
lines were detectable in our spectra. We address possible reasons
for this situation in Sect. \ref{star}.

Our data allowed us to approach the spectral line identification in
greater detail. We detected a number of previously unreported lines
of singly ionized and neutral metals. The neutral metal lines, which
are assumed to form in a disk, are single-peaked and narrow. Their
radial velocities are close to those of Fe {\sc II} lines, providing
support to our suggestion that they represent the systemic velocity.
This indicates a close to pole-on disk orientation to the line of
sight.

\subsection{Near-IR spectroscopy}\label{niris}

Our flux-calibrated near-IR spectrum of V645\,Cyg is consistent with
earlier published observations (e.g., Geballe \& Persson
\cite{gp87}, C06). It exhibits hydrogen, He {\sc I}, Fe {\sc II}, O
{\sc I} lines in emission. The He {\sc I} line at $\lambda$1.083
$\mu$m has a P Cyg type profile. Other He {\sc I} lines
($\lambda$1.701 $\mu$m and $\lambda$2.058 $\mu$m) are not clearly
detected due to a higher noise level.

The fluxes integrated within the $JHK$-bands are $\sim$ 0.8 mag
brighter than more recently obtained data (e.g., 2MASS), but are of
lower brightness than in the late 1970's (see discussion in Sect.
\ref{var}). The flux level in the optical region is lower than
expected from the published photometry. This is probably due to a
less confident removal of the atmospheric extinction, which is
higher at shorter wavelengths. Nevertheless, the spectrum clearly
shows a change in the slope at $\lambda \sim 1 \mu$m due to the
increasing contribution of the CS dust radiation.

\subsection{Speckle interferometry}\label{speckle}

Figure \ref{f1} shows the power spectra and visibility curves for
V645\,Cyg in both $H$- and $K$-bands. The azimuthally averaged
visibilities were fitted with both one- and two-component models.
One-component uniform disk and Gaussian models (the results are
shown in the middle panels of Fig. \ref{f1}) do not show good fits
in the entire range of detected visibilities, while the
two-component ring fits are more consistent with the data. The
results for two-component Gaussian and ring functions are presented
in Tables \ref{t3} and \ref{t4}, respectively. The thickness of each
ring was assumed to be 20 per cent of the ring diameter. The smaller
component contributes much more (a factor of $\sim$10) to the total
emission than the larger one in both filters. Also, both smaller and
larger components have the same size within the uncertainties in
both filters. Our data show that V645\,Cyg is clearly resolved in
both the $H$- and $K$-bands.

\section{Discussion}\label{discussion}

We now use our results along with the data collected from the
literature to place new constraints on the physical parameters of the
underlying star and its complex CS environments.

\subsection{Distance}\label{distance}

The luminosity of V645\,Cyg is not well-constrained mainly because
of its unknown distance. Some distance estimates listed in
Table \ref{t1} range from $\sim$3 to $\sim$6 kpc, while the most
recent one is 5.7$\pm$1.0 kpc (based on the CO velocity in the
direction toward the object, C06). The high IS reddening toward
V645\,Cyg is consistent with such a large distance and places the
object in the Perseus arm or even at a greater distance. To constrain
the distance more accurately, we use a method of Foster \& MacWilliams
(\cite{fm06}). They employ the non-linear modeling of the IS H {\sc I}
density along the line of sight to reconstruct the radial velocity curve.
As an example, their paper contains a radial velocity profile in the
direction ($l = 94\fdg5, b = -1\fdg5$) that is close to that of
V645\,Cyg ($l = 94\fdg6, b = -1\fdg8$). The main feature of the
velocity fields in the Perseus arm is a sharp discontinuity due to
effects of a density wave in a warped Galactic disk (described as
shocks by Foster \& MacWilliams \cite{fm06}; see their figures
3--7).

The average heliocentric radial velocity determined from the
peak positions of 37 unblended Fe {\sc II} emission lines in the
CFHT spectrum of V645\,Cyg (these lines are believed to represent
the systemic velocity, Humphreys et al. \cite{h89}) is $-44.6\pm3.5$
km\,s$^{-1}$. The H {\sc I} velocity in the line of sight toward
V645\,Cyg gradually increases with distance towards negative values
to 3.9 kpc, where it sharply decreases from $-$34 to $-$51 km\,s$^{-1}$.
At 4.5 kpc the velocity increases to $-$40 km\,s$^{-1}$ and
continues to change towards more negative velocities further away. The
object's radial velocity is reached at three different distances of
3.9, 4.3, and 5.0 kpc.

In order to resolve the distance ambiguity, complementary data are required.
A general suggestion for young objects, which have probably
not reached a state of dynamical equilibrium in Galactic motion, and
therefore the shock distance is preferred for them, is given by
Foster \& MacWilliams (\cite{fm06}). This argument ensures that the
farthest distance is unlikely. More hints can be found in our optical
spectrum. The triple IS component structure of the Na {\sc I} D--lines
(see Fig. \ref{f6}) and the K {\sc I} $\lambda$7664 \AA\
(contaminated by the telluric spectrum and not shown in Table
\ref{t5}) and $\lambda$7699 \AA\ lines can be interpreted as being due to
the absorption in the local arm (the lowest velocity component), in
the Perseus arm before the shock location (middle component), and in
the shock (the highest velocity component). Since the highest
velocity component is observed at nearly $-$50 km\,s$^{-1}$ and it
is not weak, the closest distance seems to be least favorable.
Therefore, taking into account the uncertainty in our radial
velocity measurement, one can constrain the object's distance to be
4.2$\pm$0.2 kpc. Independent estimates are still needed to verify
this result.

\subsection{Effective temperature and luminosity}\label{star}

Our high-resolution optical spectrum allows a new assessment of the
star's T$_{\rm eff}$. The absence of high excitation lines, such as
[O {\sc III}] at $\lambda$4959 and $\lambda$5007 \AA\ or [S {\sc
III}] at 6310 \AA, together with very weak emission components of
the P\,Cyg type profiles of the He {\sc I} lines suggest that
T$_{\rm eff}$ is unlikely to be higher than $\sim$25 000 K (spectral
type B1--2). Comparing the near-IR spectrum of V645\,Cyg to those of
hot stars at $\lambda \sim 2 \mu$m, C06 suggested that the object is
probably a late O--type supergiant (He {\sc I} lines in absorption, no
He {\sc II} emission lines). This comparison is inconclusive, because
line emission at $\lambda \ge$ 1.5 $\mu$m in the
spectrum of V645\,Cyg is strongly veiled by a dusty continuum. An
O-type star would give rise to He {\sc II} emission lines at
$\lambda$4686 and $\lambda$5411 \AA, which are not detected in our
spectra. Furthermore, a spectrum of \object{P\,Cyg}, a B2-type
supergiant with T$_{\rm eff} \sim$20000 K, shown in the same atlas
(Hanson, Conte, \& Rieke \cite{hcr03}) exhibits no He {\sc II} lines
and He {\sc I} and Br$\gamma$ in emission.

Pre-main-sequence stars in this temperature range are of far lower
luminosity than that estimated for V645\,Cyg (L $\sim$ (2--7)
$\times$10$^4$ L$_{\odot}$, C06). A supergiant with this temperature
and luminosity would be located well above the birthline for Herbig
Ae/Be stars (Palla \& Stahler \cite{ps93}). On the other hand,
V645\,Cyg is definitely not of an A-type. Evolved A-type supergiants
do not show such a strong emission-line spectrum, unless they belong
to the LBV group. However, even at this evolutionary stage their
spectra contain lines of a much lower excitation.

Before we proceed in constraining the stellar parameters, we
consider the suggestion by C06 that the object's apparent spectral
type represents a pseudo-photosphere within which a hotter star is
hiding. Using their upper limit to the total IR luminosity
(7$\times$10$^4$ L$_{\odot}$) and T$_{\rm eff}$ = 35000 K
appropriate to an O8 type star, we can calculate the star's radius
of $R_{\star} \sim$7 R$_{\odot}$. This is lower than a typical
main-sequence radius for such a temperature. Even the highest
luminosity quoted for V645\,Cyg in the literature (1.3$\times$10$^5$
L$_{\odot}$, Testi, Palla, \& Natta \cite{tpn98}) would only
increase $R_{\star}$ by $\sim$40 per cent and is still comparable to
those of main-sequence stars. On the other hand, the IR luminosity,
which is supposedly produced by the stellar UV radiation, does not
account for the total luminosity of the star.

\begin{figure}
\centering \resizebox{10cm}{!}{\includegraphics{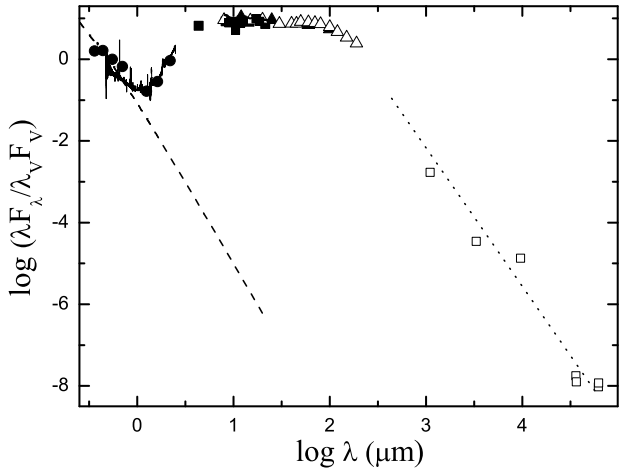}}
\caption{The SED of V645 Cyg. Optical and near-IR ground-based
photometric data are shown by filled circles, our calibrated
low-resolution spectrum by the solid line, the ISO fluxes at
selected wavelengths by open triangles, MSX data by filled squares,
IRAS fluxes by filled triangles, and millimetric and radio fluxes by
open squares. The SED is dereddened with E$(B-V)$=1.3 mag and a
standard IS reddening law from Savage \& Mathis (\cite{sm79}). The
dashed line represents a Kurucz (\cite{k94}) model atmosphere for
T$_{\rm eff}$ = 25 000 K and $\log g = 3.5$. The dotted line through
the long-wavelength part of the SED shows a blackbody fit to the
data. The fluxes are normalized to that in the $V$-band, wavelengths
are given in microns.} \label{f7}
\end{figure}

Although the amount of extinction toward the star is not well
constrained (there can be some contribution from the CS medium,
see Sect. \ref{dust}), the total flux integrated over the entire
observed spectral range (from $\lambda$0.36 $\mu$m to $\lambda$3.6
cm; see Fig. \ref{f7}) is almost extinction independent. This is
because over 90 per cent is in the IR excess, where extinction
is low. The average total dereddened observed flux is
(1.06$\pm0.03)\times 10^{-7}$ erg\,s\,cm$^{-2}$ for a range of
extinction A$_V$ from 3.1 to 4.4 mag (E$(B-V)$ from 1.0 to 1.4 mag).
At the new distance of 4.2 kpc, it corresponds to a luminosity L =
($5.8\pm0.2)\times 10^4$ L$_{\odot}$, while it would relate to L =
$(1.20\pm0.03)\times10^5$ L$_{\odot}$ at $D$ = 6 kpc (see Table
\ref{t1} and C06). In our estimate, we used the average optical
brightness from the Maidanak data, the most recent near-IR data (see
Sect. \ref{var}), ISO spectra, and sub-mm and radio data (Cohen \cite{c77};
Skinner, Brown, \& Stewart \cite{sbs93}; Di Francesco et al. \cite{df97};
Corcoran \& Ray \cite{cr98}).

The SED at $\lambda \ge$ 1 mm follows a blackbody law, which is indicative of
significant contribution from free-free emission of the outflow.
Using formulae from Panagia \& Felli (\cite{pf75}), one can estimate
that a mass-loss rate of \.M $\sim 10^{-5}$ M$_{\odot}$\,yr$^{-1}$
is required to reach the observed flux level at $\lambda$ = 3 cm
(F$_{\nu} \sim$ 0.5 mJy). This is definitely too high for an early
B-type (and even for an O-type) star. Additionally, the strong
emission from a nearly pole-on dusty disk veils the emission from
the outflow.

With the new distance, we can also estimate the luminosity of
V645\,Cyg using the average optical brightness ($ \overline{V} =
13.5\pm$0.3 mag; see Fig. \ref{f7}), the aforementioned T$_{\rm
eff}$ = 25 000 K and IS reddening. For a standard IS extinction law
(Savage \& Mathis \cite{sm79}), we find A$_V$ = 4.1 mag from the
color-excess $E(B-V)$ = 1.3 mag. Neglecting any CS effects, we derive a
visual absolute magnitude of $M_{V} = -3.7\pm0.4$. This results in a
luminosity of $\log$ L/L$_{\odot}$ = 4.4$\pm$0.1 (L =
($2.5\pm0.3)\times 10^4$ L$_{\odot}$), using an appropriate
bolometric correction of BC = $-$2.5 mag (Miroshnichenko
\cite{m98}), and a $R_{\star} \sim 8$ R$_{\odot}$.

There is a difference corresponding to a factor of two between the above
luminosity estimates, which can be reconciled in a number of ways. One can
increase T$_{\rm eff}$ to over 30 000 K to increase the bolometric
correction, although the absence of high excitation line emission would be
inconsistent with this adjustment. A higher total-to-selective IS
extinction ratio, understood to be more appropriate for
molecular clouds, could be used. For example, if we assumed that
$A_{V}/E(B-V)$ = 4 (as Natta et al. \cite{npbe93} employed) and fixed
the distance and T$_{\rm eff}$ values, the luminosity would match
the one derived from the integrated flux. However, since most of the
IS extinction originates in the foreground of the object's molecular
cloud, it is more appropriate to use an average Galactic ratio for
hot stars of 3.1. Additionally, the integrated flux may contain a
significant contribution from light scattered off the CS dust
that would not otherwise be directed along the line of sight.
Whitney et al. (\cite{wwbw03}) demonstrated that the integrated flux of a
star with a pole-on disk may exceed that of the star alone by a
factor of up to two. Taking this effect into account would decrease
the difference between the two luminosity estimates.

The possible contribution of free-bound and free-free emission from
the CS gas to the continuum may also affect the luminosity estimate
derived from the optical brightness. Thus, it is likely that the true
luminosity of the underlying star is in between our two estimates.
With all possible sources of uncertainty, the above estimates of the
star's parameters appear reasonable and place its position as being
close to the main-sequence, where a massive young star that has already
appeared from its protostellar cocoon is located.

Nevertheless, there remains the problem of not detecting photospheric
lines in the object's optical spectrum. Since the CS disk is
oriented nearly pole-on (see Sect. \ref{dust}), one can expect
the star's rotation axis to be oriented nearly along the line of sight.
Therefore, rotational line broadening would not have an important
impact on the photospheric line profiles. Some lines might have been
missed in our spectrum because of its low S/N ratio, especially in
the blue region. Some other lines (e.g., H {\sc I}, He {\sc I}, Si
{\sc II}) are entirely due to the CS gas, so that no photospheric
component can be detected. Thus, the most likely reason for this
situation is the aforementioned CS continuum radiation. This must be
extremely strong, since a number of other objects with even more distinctive
emission-line spectrum still exhibit photospheric lines at similar
spectral resolution and S/N ratio (e.g., Miroshnichenko et al.
\cite{m04}). However, we were unable to detect reliably any photospheric lines
even in regions relatively free of line emission (e.g., $\lambda\lambda$
5600--5800 \AA), of high S/N ratios ($\ga$70) in our spectra, and
where some photospheric lines for the derived spectral type are
expected (e.g., S {\sc II} $\lambda$5640 \AA\ or N {\sc II}
$\lambda$5680 \AA). Future observations of even higher
S/N ratios are definitely required to address this problem further.

\subsection{Brightness variations}\label{var}

\begin{figure}
\centering \resizebox{10cm}{!}{\includegraphics{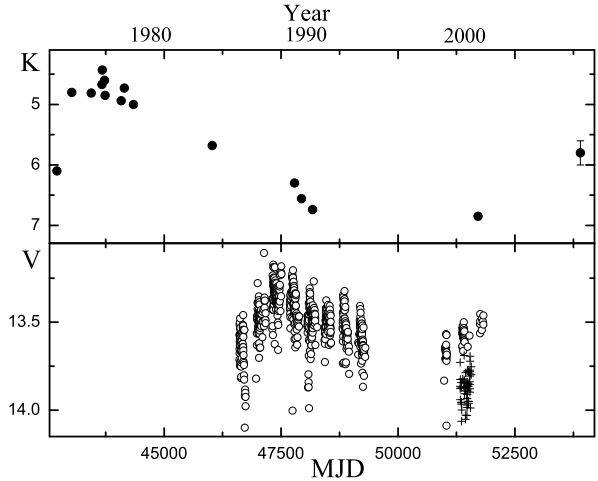}}
\caption{Optical and near-IR light curves of V645\,Cyg. The
observations obtained at the Maidanak observatory and published by
Shevchenko et al. (\cite{sh93}) are shown by open circles. The
observations from the NSVS survey (Wozniak et al. \cite{w04}) and
color-corrected using the mean $B-V$ color from the Maidanak data
are shown by pluses. The $K$-band data are collected from the
literature and described in the text. The last data point (2006) on
the $K$-band light curve represents our near-IR spectrum. The time
is shown in modified Julian dates MJD = JD$-$2400000.} \label{f8}
\end{figure}

Long-term monitoring of V645\,Cyg was performed at the Maidanak
Observatory in Uzbekistan between 1985 and 2000 (740 $UBVR$
observations, progress report published by Shevchenko et al.
\cite{sh93}). The $V$-band light curve is shown in the lower panel
of Fig. \ref{f8}. Both knots, N0 and N1, were included in the
aperture of these observations. It confirms the old result of
Hoffmeister et al. (\cite{hra51}) about the long-term slow
brightness variations. A brightness modulation with a possible
period of $\sim$8 years and an amplitude of $\sim$0.6 mag is seen in
the Maidanak data. The color-indices are stable to within $\sim$0.1 mag
and resemble those of a reddened early B-type star ($U-B$=0.2 mag,
$B-V$=1.1 mag, $V-R$=1.2 mag).

A number of near-IR photometric observations of V645\,Cyg have been
obtained since its discovery as an IR source in the 1970's. However,
the entire collection has not yet been studied. A careful look at
the data shows a large scatter in the object's brightness in the
1--3 $\mu$m region (see the upper panel of Fig. \ref{f8}). The first
published IR photometry by Lebofsky et al. (\cite{l76}) was obtained
on 1975 October 24 and resulted in a $K$-band flux of 2.3 Jy, which
corresponds to 6.1 mag (using a standard calibration of 653 Jy for a
zero-point of the magnitude scale, Cohen et al. \cite{c92}). The
accuracy of the observational data was not indicated in the original paper.
Several other papers (Gosnell, Hudson, \& Puetter \cite{g79};
Humphreys et al. \cite{hmb80}; Grasdalen et al. \cite{g83})
reporting IR photometry of V645\,Cyg obtained in 1976--1980 detected
a variable $K$-band brightness in a range of 4.4--5.0 mag. with a
slight tendency of fading with time. This brightness level has usually
been adopted in studies of the object's SED. All of these observations were
obtained with single-element detectors with relatively large
apertures (typically over 10$\arcsec$). Humphreys et al.
(\cite{hmb80}) noted that there was no difference in the detected
signal from aperture size between 9$\arcsec$ and 26$\arcsec$. This
indicates that the N1 knot was probably always present within the
apertures.

The next published near-IR observation obtained in 1984 or 1985
(Geballe \& Persson \cite{gp87}) was a spectrum, taken through a
5$\arcsec$ diameter aperture with a continuum flux of
$\sim$2.2$\times$10$^{-16}$ W\,cm$^{-2}$\,$\mu$m at $\lambda$2.3
$\mu$m, which roughly translates into a $K$-band brightness of 5.7
mag. It excluded the N1 knot and only provided information about the
continuum brightness of the N0 condensation. Although this
measurement could not be properly compared with earlier data, it
appears to be the first indication of a significant decline in the
object's near-IR brightness. All photometric observations published
subsequently that used both single- and multiple-element
detectors (Minchin et al. \cite{mhby91}; Sun et al. \cite{s91};
Noguchi et al. \cite{n93}; Skrutskie et al. \cite{s06}) measured
even lower $K$-band brightness of 6.3--6.8 mag. The flux level in
our near-IR spectrum integrated over the $K$-band response curve
implied a brightness of 5.8$\pm$0.2 mag in agreement with the data obtained
after 1984.

Studying the published near-IR data for V645\,Cyg, we noticed that
the relative brightness of the main source (N0) and the N1 knot
seemed to change with time. They had almost the same brightness in a
$K$-band image taken by Hodapp (\cite{h94}) in 1990, and a 2MASS image
taken in 2000 indicated that N0 was nearly twice as bright as N1, while
C06 claimed that N0 was $\sim$ 200 times brighter than N1 in their
2002 image (although the center of N0 is saturated in this
image). Although this temporal trend still has to be
confirmed, two important implications can be still deduced. First,
the decrease in near-IR brightness of V645\,Cyg cannot be entirely due
to the fading of N1 alone. Second, simultaneous fading of both N0 and N1
may be due to a decrease in the amount of reflective matter close to the
star. This can in turn be a consequence of the strong outflow that
gradually removes the protostellar cloud material from the immediate
vicinity of the star.

The sparseness of the near-IR data and a short-time overlap with the optical
data do not permit a clear comparison with the object's behaviour at
different wavelengths, although there is some indication of a
positive correlation based on the five near-IR data points after
1990. If this is the case, then the brightness variations can be
explained by luminosity fluctuations of the illumination source or
changes in the CS gas properties. The latter is consistent with no
or little CS dust being present along the line of sight (see Sect.
\ref{dust}).

\begin{figure}
\centering \resizebox{10cm}{!}{\includegraphics{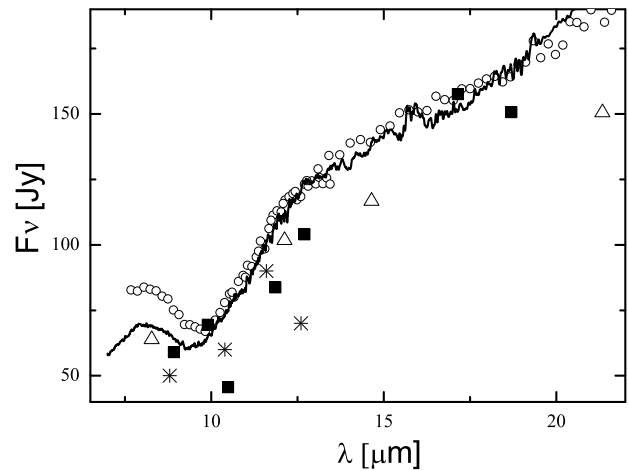}}
\caption{Mid-IR part of the SED of V645\,Cyg. Symbols: solid line --
ISO SWS data, open circles -- IRAS LRS data, triangles -- MSX data,
filled squares -- ground-based photometry from Salas,
Gruz-Gonz\'ales, \& Tapia \cite{sgt06}), asterisks -- ground-based
photometry from Lebofsky et al. (\cite{l76}). The observed scatter
is most likely due to uncertainties in absolute calibration.}
\label{f9}
\end{figure}

There has been only minor variation in the $H-K$ color-index, which averages
to be 2.3$\pm$0.1 mag over all 30 years of observations. The mid-IR
part of the SED has also been stable (see Fig. \ref{f9}). Therefore, it
appears that no significant changes in the CS dust structure has
occurred.

\subsection{Outflow properties}\label{outflow}

The presence of many P Cyg type profiles in the spectrum of
V645\,Cyg indicates that the underlying star drives a strong
outflow. Another feature of the spectrum is forbidden lines (see
Fig. \ref{f5}), which are seen only blueward of the systemic radial
velocity. This rare situation was interpreted as the blocking of the
receding part of the outflow by an optically-thick disk, where
low-excitation lines form (e.g., K {\sc I}, Fe {\sc I}, and Ca {\sc
II}; Hamann \& Persson \cite{hp89}). It also prompted a search for
the possible signs of the receding part. C06 suggested that the
He {\sc I} 1.701 $\mu$m line and two emission features (with a
radial velocity of +2000 km\,s$^{-1}$ with respect to the center
of the H$\alpha$ line ($\lambda$6604 \AA\ and with a radial velocity
of +1800 km\,s$^{-1}$ with respect to the CO 2.29 $\mu$m band head)
might represent the receding region.

We identified the $\lambda$6604 \AA\ feature as a Sc {\sc II}
line, but the quality of our near-IR spectrum was not high enough to
assess the origin of the remaining two. Nevertheless, we may ask whether
the receding outflow has truly been missed in the hydrogen, helium, and
ionized iron lines? Emission components of these line profiles are smooth, showing
no evidence of any cutoff, as one would expect from an obscuration. The CS material
density decreases outward in the accelerated outflow (see Hamann \&
Persson \cite{hp89} for discussion), such that forbidden lines, which
are of highest intensity in regions of relatively low density, form much
further away from the star, beyond the region where the outflow reaches
its terminal velocity of $\sim$800 km\,s$^{-1}$ (see Fig. \ref{f3}).
The rapidly moving material may decelerate there, colliding with material
from the ambient molecular cloud, which would explain the lower velocities
of the forbidden lines. A possible deceleration of the molecular outflow from
V645\,Cyg was suggested by Verdes-Montenegro et al. (\cite{vm91}).
On the other hand, the initial acceleration cannot have been incredibly high,
because the density would have had to decline rapidly, leading to weak line
emission. Therefore, it is reasonable to conclude that the inner edge of the
optically-thick disk is located at a distance of at least at a few tens of
the stellar radii.

\subsection{Structure of the CS dust}\label{dust}

Clearly, V645\,Cyg is surrounded by a large amount of CS dust that
manifests itself in a strong IR excess (see Fig. \ref{f7}). The
absence of significant obscuration toward the star (the reddening
deduced from optical photometry is similar to that inferred from
the DIB strengths; see Sect. \ref{optical}), along with the presence
of a disk-like structure, which blocks the red parts of the
forbidden lines, indicate that the CS dust geometry may also be
disk-like. However, it has already been suggested that V645\,Cyg may
have an optically-thin dusty envelope in addition to a
geometrically-thin and optically-thick (at least at short
wavelengths) disk (Natta et al. \cite{npbe93}). This suggestion is
supported by measurements of the object's IR size, which increases
with wavelength in the near- and mid-IR ($\le 14\arcsec$ at $\lambda
50 \mu$m and 22$\arcsec \pm 2\arcsec$ at $\lambda 100 \mu$m, Natta
et al. \cite{npbe93}) and decreases at yet longer wavelengths ($\sim
14\arcsec$ at $\lambda 800 \mu$m, Sandell \& Weintraub \cite{sw94},
$\sim 2\arcsec$ at $\lambda$ 3.6 cm, Di Francesco et al.
\cite{df97}). Miroshnichenko et al. (\cite{m99}) demonstrated that no
one-component source can reproduce this size variation with
wavelength. For the flux integrated over the entire object, a colder
and smaller disk dominates the longer-wavelength domain ($\lambda
\ge 100 \mu$m), while a warmer and more extended envelope is more dominant
at shorter wavelengths.

Another important question that has not been clearly addressed is
whether accretion is still ongoing or not in the disk. Natta et al.
(\cite{npbe93}) argued that the disk dominates the energy output of
V645\,Cyg and its contribution is much larger than expected
for a passive reprocessing disk. They estimated the accretion rate to
be $7\,\times 10^{-3}$ M$_{\odot}$\,yr$^{-1}$, which is much
higher than a typical protostellar accretion rate for
intermediate-mass stars of $10^{-5} - 10^{-4}$
M$_{\odot}$\,yr$^{-1}$ (Palla \& Stahler \cite{ps93}). Since the
star is probably more massive than 10 M$_{\odot}$, intense
accretion should have halted while it was still inside its parental
cloud. Since the star has already been released from the cocoon, it
must be already on the main-sequence. The outflow properties
(the fact that we see no long-wavelength cutoff of the lines that form
close to the star, see Sect. \ref{outflow}) also imply that the disk
does not approach close to the star. These arguments strongly suggest
that accretion is currently absent, and the star is the main source of
energy in the system.

\subsection{Speckle interferometry of the dusty environments}\label{dust1}

The location of the $H$-band close to the minimum of the
dereddened SED (see Fig. \ref{f7}) may imply that the systematically
larger size (although within the uncertainties) of the best-fit model in
comparison with that of the $K$-band can be due to a higher
scattered light contribution. Thus, we use the fitting results
for the $K$-band to characterize the distance of the hottest dust
from the star.

Using the theory of radiative transfer in dusty media, Ivezi\'c \&
Elitzur (\cite{ie97}) showed that, irrespective of the envelope
geometry, the distance of the closest dust to the radiation source,
$r_{1}$, is determined by the source luminosity and a scaling factor
$\Psi$, which depends on the dimensionless profiles of the source
radiation and the dust absorption efficiency, such that

\begin{equation}\label{eq1}
    {r_{1} \over R_{\star}} = {\sqrt{\Psi} \over 2}~ \left({T_{\rm eff} \over
    T_{1}}\right)^2,
\end{equation}

\noindent where $T_1$ is the dust temperature at $r_1$.

They denote by $\Psi$ a ratio of the wavelength dependence of the dust
absorption efficiency weighted with the radiation source SED, to the
identical quantity weighted by the dust radiation (blackbody) SED for $T_1$.
This factor was calculated for a typical $T_1$ of 1500 K by Monnier
\& Millan-Gabet (\cite{mm02}) for a range of T$_{\rm eff}$ and dust
optical properties. They showed that the factor is almost unity for
large particles (1 $\mu$m size) and sharply increases to a
factor of 50--80 for grain sizes of the order of 0.01 $\mu$m.

Our most suitable single-component uniform disk fit with a diameter of
37.0$\pm$8.5 mas corresponds to a distance of 78$\pm$19 AU at 4.2
kpc or 1660$\pm$400 $R_{\star}$, assuming that $R_{\star}$ = 10
$R_{\odot}$. This requires $\Psi \sim$140 to have the closest dust
at such a large distance from the star. The result is inconsistent
with even the smallest dusty grains. These grains, if they exist in the
envelope or the disk surface layer of V645\,Cyg, would produce
emission features (e.g., at $\lambda 9.7 \mu$m) that are not
observed in its spectrum (see Fig. \ref{f9}).

\begin{figure}
\centering \resizebox{10cm}{!}{\includegraphics{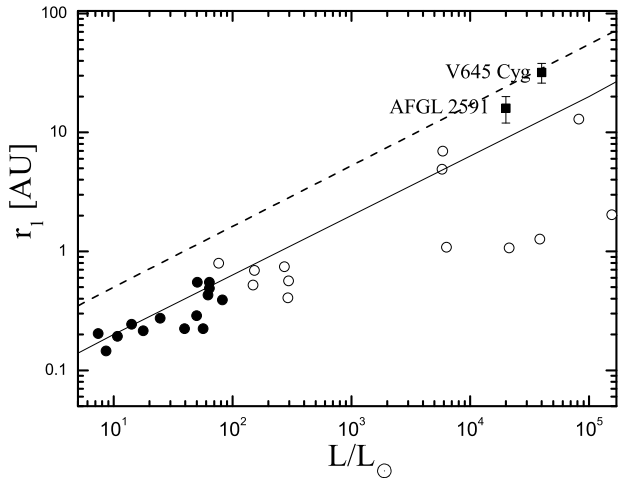}}
\caption{Measured sizes of Herbig Ae (filled circles) and Herbig Be
(open circles) stars plotted versus their luminosity (adopted from
Millan-Gabet et al. \cite{mg07}). Data for AFGL\,2591 are taken from
Preibisch et al. (\cite{pr03}), data for V645\,Cyg are from this
work. The lines show the expected dust sublimation radii for
temperatures of 1000 K (dashed line) and 1500 K (solid line), warmed
by direct heating with no backwarming.} \label{f10}
\end{figure}

Therefore, it seems more appropriate to use a two-component fit.
Since fits to interferometric data for objects with CS
envelopes commonly assume a ring-like matter distribution (a typical
width of such a ring is 20 per cent of its diameter; see Monnier et
al. \cite{mon05}). This simple method allows us to approximate a rough
characteristic size that can be used for comparison with other
observations. The best-fit model in the $K$-band has a radius of the
compact component of 12$\pm$2 mas that translates into a distance of
26$\pm$4 AU or 510$\pm$85 $R_{\star}$. This requires a value of
$\Psi \sim$10, and corresponds to an average grain size of $\sim$0.3
$\mu$m. Another estimate can be obtained by using the calculations of dust
temperatures that would produce the observed size-luminosity
relationship presented by Millan-Gabet et al. (\cite{mg07}). They
are shown in Fig. \ref{f10} and suggest that the size of V645\,Cyg
at $\lambda 2\mu$m can be explained by radiation from 1000 K dust.
Taking into account the uncertainties in the star's temperature and
the dust distribution, we can conclude that the small component of
the $K$-band visibility fit represents the CS dust close to its
sublimation radius. It cannot be excluded that the inner dust
boundary has already been moved by the strong outflow from the star. The
lack of attenuation by the CS dust is in agreement with the image
azimuthal symmetry.

The larger component of the fit represents a much larger structure
and is also much fainter. Its radius in the $K$-band corresponds to
251$\pm$22 AU or $\sim 5\times10^{3} R_{\star}$. It is reasonable to
suggest that the smaller component represents the dust sublimation
distance, while the larger one represents the dust at $\sim$9 $r_1$.
On small scales, the disk's inner rim is brighter than the
optically-thin envelope, which dominates the spatially integrated
flux from the object in the $K$-band (see Sect. \ref{dust}).
Therefore, we can use the calculations of Ivezi\'c \& Elitzur
(\cite{ie97}) for such a regime that exhibit a temperature decrease of
a factor of three from $T_1$ at this distance. Thus, the larger
component can be due to the inner part of the optically thin and
tenuous CS dust envelope, which can be driven away by the outflow. A
contribution from the light reflected off the dusty disk or outflow
cavity cannot be excluded.

The angular size of V645\,Cyg in the $K$-band is comparable to that
inferred from similar speckle interferometric observations of
another young massive stellar object \object{AFGL\,2591} (Preibisch
et al. \cite{pr03}). The central star in AFGL\,2591 is invisible due
to obscuration by the CS matter, so that the inner rim of the dusty
disk illuminated by stellar radiation is observed. However, these
two objects have some features in common. First, their
single-telescope speckle interferometric near-IR data can be more accurately
fitted with a two-component model. Second, even the smaller
component of these fits corresponds to linear sizes, which are the
largest among those observed for young stars (see Fig. \ref{f10}).
They also have almost the highest luminosities of all observed
objects, which agrees with expectations for the sizes. Both
objects are currently understood to be massive young stars and
certainly require careful monitoring to be able to follow any possible rapid
evolution in their CS environments (e.g., as described in Sect. \ref{var}).

\section{Conclusions}\label{conclus}

The new spectroscopic and speckle interferometric observations of
V645\,Cyg have resulted in new constraints on the object's stellar and CS
parameters. The object has been resolved at near-IR wavelengths for the
first time. We found that photospheric features may not be evident and
detectable even at high spectral resolution and relatively high S/N ratio.
Combination of the existing photometric and spectral data appear to
exclude the A-type classification of V645\,Cyg. The star is also
unlikely to be of O-type. The line excitation level in the
optical spectrum indicates that the star is within the early B-type
domain (T$_{\rm eff} \sim$ 25 000 K).

We have used a new method of distance determination based on the H {\sc
I} density profile along the line of sight (Foster \& Macwilliams
\cite{fm06}) to constrain the object's location in the Galaxy. The
average radial velocity of unblended Fe {\sc II} lines along with
the structure of the IS lines suggest a new distance of 4.2$\pm$0.2
kpc toward V645\,Cyg. At this distance, the star's luminosity is most
likely to be between 2$\times$10$^4$ and 6$\times$10$^4$ L$_{\odot}$,
still in agreement with many previous estimates.

Spectral line profiles in the optical region are consistent with the
presence of an opaque disk that blocks the region of the outflow
where forbidden lines form, but does not affect the very inner parts
where hydrogen and singly ionized metal lines originate. The CS dust is
located in an optically-thick disk, which is oriented almost pole-on
and dominates the radiation at $\lambda \ge 100 \mu$m, and in an
optically-thin tenuous envelope, which is responsible for most of
the near-IR and mid-IR radiation. This kind of distribution is
consistent with the measurements of the object's size in the IR
region that increases with wavelength up to $\lambda \sim 100 \mu$m
and decreases at longer wavelengths.

Summarizing the above analysis, we suggest that V645\,Cyg is an
early B-type star that recently appeared from its molecular cloud
and started its main-sequence evolution. These objects are rare,
because they are rather massive and evolve fast. The orientation of the
CS disk ensures that V645\,Cyg is unique, since many objects still surrounded
by the protostellar-cloud remains are attenuated by their CS disks.

Reviewing the available data, we have revealed several interesting
results that have not been reported earlier. In particular, the near-IR
brightness of V645\,Cyg has decreased by nearly 2 mag since the 1980's.
However, our absolutely calibrated near-IR spectrum obtained in 2006 indicates
that this process may be temporary.
Also, the relative brightness of the N1 knot with respect to that of the star
plus disk/envelope appears to have decreased significantly in the same
spectral region. A correlation between the variations in the optical
and near-IR brightness was suspected. We strongly recommend
follow-up photometric and spectral monitoring of V645\,Cyg.
Photometry can easily be obtained at small telescopes, while
spectroscopy, necessarily of high-resolution to search for
photospheric lines and emission-line-profile variations, requires
larger facilities.

\begin{acknowledgements}
A.M. thanks the Max-Planck-Society for support during his stay in
Bonn. We thank the anonymous referee for the suggestions that
allowed us to improve the data presentation and analysis. This
publication makes use of the SIMBAD database operated at CDS,
Strasbourg, France, the NSVS data base, and data products from the
Two Micron All Sky Survey, which is a joint project of the
University of Massachusetts and the Infrared Processing and Analysis
Center/California Institute of Technology, funded by the National
Aeronautics and Space Administration and the National Science
Foundation.
\end{acknowledgements}

\begin{table*}
\caption{Line identification in the optical spectrum of V645\,Cyg}
\label{t5} \centering {\tiny
\begin{tabular}{crrrlclcrrrlcl}
\hline\noalign{\smallskip}
$\lambda$& EW    & I/I$_c$& V$_r$& ID      & FWHM  & Comment &$\lambda$& EW    & I/I$_c$& V$_r$& ID      & FWHM  & Comment \\
\noalign{\smallskip}\hline\noalign{\smallskip}
 4320.97 & 0.36 &$\!\!\!$ 1.33 &$\!\!\!$$-$62.5& Ti II  (41)&$\!\!\!$  1.02 &  n               & 5657.92 &  0.69 &$\!\!\!$ 1.32 &$\!\!\!$$-$38.2& Fe II (57)&$\!\!\!$  2.14 &  b        \\
 4330.26 & 0.00 &$\!\!\!$ 1.33 &$\!\!\!$$-$47.8& Ti II  (94)&$\!\!\!$  0.00 &  n, b w/H$\gamma$& 5662.94 &  0.28 &$\!\!\!$ 1.14 &$\!\!\!$$-$47.7& Fe  I (924)&$\!\!\!$ 1.84 &  ID:      \\
 4340.47 & 3.20 &$\!\!\!$ 0.30 &$\!\!\!$$-$313.1& H   I  ( 1)&$\!\!\!$  5.31 &  P Cyg          & 5667.16 &  0.00 &$\!\!\!$ 1.11 &$\!\!\!$$-$33.9& Sc II  (29)&$\!\!\!$ 0.00 &  +f       \\
 4340.47 & 7.18 &$\!\!\!$ 3.08 &$\!\!\!$$-$12.5&  H   I  ( 1)&$\!\!\!$  4.12 &                 & 5669.03 &  0.42 &$\!\!\!$ 1.15 &$\!\!\!$$-$48.7& Sc II  (29)&$\!\!\!$ 0.00 &           \\
 4351.76 & 2.18 &$\!\!\!$ 1.86 &$\!\!\!$$-$45.5& Fe II  (27)&$\!\!\!$  2.58 &                  & 5684.19 &  0.22 &$\!\!\!$ 1.14 &$\!\!\!$$-$42.8& Sc II  (29)&$\!\!\!$ 1.59 &           \\
 4374.83 & 0.00 &$\!\!\!$ 1.88 &$\!\!\!$$-$40.4& Ti II  (93)&$\!\!\!$  0.00 &  b               & 5780.41 &  0.63 &$\!\!\!$ 0.74 &$\!\!\!$$-$27.0& DIB        &$\!\!\!$ 2.32 &           \\
 4385.38 & 3.60 &$\!\!\!$ 2.10 &$\!\!\!$$-$52.7& Fe II  (27)&$\!\!\!$  3.45 &  n, b:           & 5796.03 &  0.20 &$\!\!\!$ 0.84 &$\!\!\!$$-$28.5& DIB        &$\!\!\!$ 1.16 &           \\
 4395.03 & 1.74 &$\!\!\!$ 1.76 &$\!\!\!$$-$34.1& Ti II  (19)&$\!\!\!$  2.13 &                  & 5889.95 &  0.00 &$\!\!\!$ 0.11 &$\!\!\!$$-$58.1& Na  I  (11)&$\!\!\!$ 0.00 &           \\
 4399.86 & 1.32 &$\!\!\!$ 1.70 &$\!\!\!$$-$40.9& Fe II  (20)&$\!\!\!$  2.17 &                  & 5889.95 &  0.00 &$\!\!\!$ 0.00 &$\!\!\!$$-$35.7& Na  I  (11)&$\!\!\!$ 0.00 &           \\
 4416.82 & 1.78 &$\!\!\!$ 1.59 &$\!\!\!$    0.7& Fe II  (27)&$\!\!\!$  3.09 &  b:              & 5889.95 &  0.00 &$\!\!\!$ 0.00 &$\!\!\!$$-$10.2& Na  I  (11)&$\!\!\!$ 0.00 &           \\
 4421.95 & 0.30 &$\!\!\!$ 1.20 &$\!\!\!$$-$23.8& Ti II  (93)&$\!\!\!$  1.61 &                  & 5895.92 &  0.00 &$\!\!\!$ 0.17 &$\!\!\!$$-$57.5& Na  I  (11)&$\!\!\!$ 0.00 &           \\
 4443.80 & 1.51 &$\!\!\!$ 1.57 &$\!\!\!$$-$37.8& Ti II  (19)&$\!\!\!$  2.50 &                  & 5895.92 &  0.00 &$\!\!\!$ 0.00 &$\!\!\!$$-$36.6& Na  I  (11)&$\!\!\!$ 0.00 &           \\
 4489.19 & 0.00 &$\!\!\!$ 1.66 &$\!\!\!$$-$44.8& Fe II  (37)&$\!\!\!$  0.00 &  +f              & 5895.92 &  0.00 &$\!\!\!$ 0.09 &$\!\!\!$$-$12.7& Na  I  (11)&$\!\!\!$ 0.00 &           \\
 4491.40 & 3.48 &$\!\!\!$ 1.88 &$\!\!\!$$-$57.4& Fe II  (37)&$\!\!\!$  4.84 &                  & 5991.38 &  0.90 &$\!\!\!$ 1.41 &$\!\!\!$$-$42.6& Fe II  (46)&$\!\!\!$ 2.16 &           \\
 4501.27 & 0.64 &$\!\!\!$ 1.42 &$\!\!\!$$-$54.6& Ti II  (31)&$\!\!\!$  1.43 &  P Cyg:          & 6084.11 &  0.45 &$\!\!\!$ 1.21 &$\!\!\!$$-$44.9& Fe II  (46)&$\!\!\!$ 2.15 &           \\
 4508.28 & 1.14 &$\!\!\!$ 1.68 &$\!\!\!$$-$47.2& Fe II  (38)&$\!\!\!$  1.71 &  P Cyg           & 6102.72 &  0.05 &$\!\!\!$ 1.03 &$\!\!\!$$-$23.6& Ca  I  ( 3)&$\!\!\!$ 1.50 &           \\
 4515.34 & 0.60 &$\!\!\!$ 1.44 &$\!\!\!$$-$48.5& Fe II  (37)&$\!\!\!$  1.37 &  P Cyg           & 6113.33 &  0.00 &$\!\!\!$ 1.05 &$\!\!\!$$-$63.3& Fe II  (46)&$\!\!\!$ 0.00 &  +f       \\
 4520.23 & 0.00 &$\!\!\!$ 1.34 &$\!\!\!$$-$37.2& Fe II  (37)&$\!\!\!$  0.00 &  +f              & 6114.60 &  0.00 &$\!\!\!$ 1.05 &$\!\!\!$$-$36.8& N  II  ( 7)&$\!\!\!$ 0.00 &           \\
 4522.63 & 1.88 &$\!\!\!$ 1.67 &$\!\!\!$$-$50.4& Fe II  (38)&$\!\!\!$  3.41 &                  & 6122.22 &  0.09 &$\!\!\!$ 1.05 &$\!\!\!$$-$49.0& Ca  I  ( 3)&$\!\!\!$ 1.65 &           \\
 4534.16 & 1.15 &$\!\!\!$ 1.54 &$\!\!\!$$-$50.9& Fe II  (37)&$\!\!\!$  1.34 &  P Cyg:          & 6129.71 &  0.09 &$\!\!\!$ 1.05 &$\!\!\!$$-$52.4& Fe II  (46)&$\!\!\!$ 1.88 &           \\
 4549.47 & 1.80 &$\!\!\!$ 1.90 &$\!\!\!$$-$38.9& Fe II  (38)&$\!\!\!$  1.95 &  P Cyg           & 6141.72 &  0.16 &$\!\!\!$ 1.12 &$\!\!\!$$-$45.9& Ba II  ( 2)&$\!\!\!$ 1.56 &           \\
 4555.89 & 1.17 &$\!\!\!$ 1.50 &$\!\!\!$$-$77.7& Fe II  (37)&$\!\!\!$  2.23 &  b:              & 6147.73 &  1.93 &$\!\!\!$ 1.54 &$\!\!\!$ $-$2.9& Fe II  (74)&$\!\!\!$ 3.28 &  b:       \\
 4558.58 & 0.54 &$\!\!\!$ 1.41 &$\!\!\!$$-$43.4& Fe II  (20)&$\!\!\!$  1.16 &                  & 6191.56 &  0.33 &$\!\!\!$ 1.18 &$\!\!\!$$-$41.2& Fe  I (169)&$\!\!\!$ 1.79 &           \\
 4563.76 & 0.82 &$\!\!\!$ 1.51 &$\!\!\!$$-$40.1& Ti II  (50)&$\!\!\!$  1.51 &  P Cyg:          & 6195.00 &  0.06 &$\!\!\!$ 0.95 &$\!\!\!$$-$14.5& DIB        &$\!\!\!$ 0.84 &           \\
 4571.97 & 0.94 &$\!\!\!$ 1.55 &$\!\!\!$$-$42.0& Ti II  (82)&$\!\!\!$  1.64 &  P Cyg           & 6219.54 &  0.14 &$\!\!\!$ 1.06 &$\!\!\!$$-$47.3& Fe II  (34)&$\!\!\!$ 2.50 &           \\
 4576.33 & 0.26 &$\!\!\!$ 1.30 &$\!\!\!$$-$48.5& Fe II  (38)&$\!\!\!$  0.77 &                  & 6238.38 &  1.35 &$\!\!\!$ 1.46 &$\!\!\!$$-$32.2& Fe II  (74)&$\!\!\!$ 2.83 &           \\
 4583.82 & 1.90 &$\!\!\!$ 1.95 &$\!\!\!$$-$46.5& Fe II  (38)&$\!\!\!$  1.91 &  P Cyg           & 6247.56 &  2.04 &$\!\!\!$ 1.72 &$\!\!\!$$-$48.5& Fe II  (74)&$\!\!\!$ 2.92 &           \\
 4588.21 & 0.00 &$\!\!\!$ 1.47 &$\!\!\!$$-$26.8& Cr II  (44)&$\!\!\!$  0.00 &  +2f             & 6252.56 &  0.11 &$\!\!\!$ 1.05 &$\!\!\!$$-$31.2& Fe  I (169)&$\!\!\!$ 2.02 &           \\
 4589.89 & 0.00 &$\!\!\!$ 1.47 &$\!\!\!$$-$56.9& Cr II  (44)&$\!\!\!$  0.00 &                  & 6300.23 &  0.00 &$\!\!\!$ 1.73 &$\!\!\!$$-$394.7& O  I  ( 1)&$\!\!\!$ 0.00 &           \\
 4592.09 & 2.15 &$\!\!\!$ 1.29 &$\!\!\!$$-$50.3& Cr II  (44)&$\!\!\!$  0.00 &                  & 6300.23 &  0.00 &$\!\!\!$ 1.77 &$\!\!\!$$-$62.4& O   I  ( 1)&$\!\!\!$ 0.00 &           \\
 4618.83 & 0.00 &$\!\!\!$ 1.43 &$\!\!\!$$-$39.6& Cr II  (44)&$\!\!\!$  0.00 &  +f              & 6318.11 &  0.25 &$\!\!\!$ 1.11 &$\!\!\!$$-$52.2& Ca  I  (53)&$\!\!\!$ 2.13 &           \\
 4620.51 & 0.00 &$\!\!\!$ 1.40 &$\!\!\!$$-$48.7& Fe II  (38)&$\!\!\!$  0.00 &                  & 6363.88 &  0.00 &$\!\!\!$ 1.31 &$\!\!\!$$-$402.6& O  I  ( 1)&$\!\!\!$ 0.00 &           \\
 4629.34 & 1.41 &$\!\!\!$ 1.78 &$\!\!\!$$-$46.6& Fe II  (38)&$\!\!\!$  1.84 &  P Cyg           & 6363.88 &  0.00 &$\!\!\!$ 1.26 &$\!\!\!$$-$72.6& O   I  ( 1)&$\!\!\!$ 0.00 &           \\
 4634.11 & 1.17 &$\!\!\!$ 1.45 &$\!\!\!$$-$40.8& Cr II  (44)&$\!\!\!$  2.73 &  P Cyg +p        & 6369.45 &  0.36 &$\!\!\!$ 1.16 &$\!\!\!$$-$34.4& Fe II  (40)&$\!\!\!$ 2.26 &  P Cyg    \\
 4648.23 & 0.00 &$\!\!\!$ 1.10 &$\!\!\!$$-$43.9& Fe II  (38)&$\!\!\!$  0.00 &  n               & 6379.00 &  0.06 &$\!\!\!$ 0.94 &$\!\!\!$$-$17.9& DIB        &$\!\!\!$ 0.75 &           \\
 4656.97 & 0.45 &$\!\!\!$ 1.22 &$\!\!\!$$-$44.5& Fe II  (43)&$\!\!\!$  1.88 &  n, P Cyg        & 6393.61 &  0.18 &$\!\!\!$ 1.10 &$\!\!\!$$-$36.6& Fe  I (168)&$\!\!\!$ 1.70 &           \\
 4666.75 & 0.40 &$\!\!\!$ 1.25 &$\!\!\!$$-$39.2& Fe II  (37)&$\!\!\!$  1.40 &                  & 6400.01 &  0.13 &$\!\!\!$ 1.05 &$\!\!\!$$-$42.2& Fe  I (816)&$\!\!\!$ 2.41 &           \\
 4670.17 & 0.41 &$\!\!\!$ 1.28 &$\!\!\!$$-$27.6& Fe II  (25)&$\!\!\!$  1.34 &                  & 6407.30 &  0.17 &$\!\!\!$ 1.06 &$\!\!\!$$-$47.3& Fe II  (74)&$\!\!\!$ 2.73 &  b:       \\
 4708.90 & 0.34 &$\!\!\!$ 1.16 &$\!\!\!$$-$51.6& Ba II  (15)&$\!\!\!$  1.96 &                  & 6416.91 &  0.92 &$\!\!\!$ 1.37 &$\!\!\!$$-$41.6& Fe II  (74)&$\!\!\!$ 2.39 &  b:       \\
 4731.44 & 0.77 &$\!\!\!$ 1.45 &$\!\!\!$$-$49.4& Fe II  (43)&$\!\!\!$  1.75 &                  & 6432.65 &  1.80 &$\!\!\!$ 1.71 &$\!\!\!$$-$43.8& Fe II  (40)&$\!\!\!$ 2.59 &           \\
 4764.54 & 0.73 &$\!\!\!$ 1.36 &$\!\!\!$$-$72.4& Ti II  (48)&$\!\!\!$  2.10 &  b               & 6456.38 &  0.43 &$\!\!\!$ 0.93 &$\!\!\!$$-$328.5&Fe II  (74)&$\!\!\!$ 5.62 &           \\
 4779.99 & 0.56 &$\!\!\!$ 1.33 &$\!\!\!$$-$41.4& Ti II  (92)&$\!\!\!$  1.85 &                  & 6456.38 &  2.19 &$\!\!\!$ 1.83 &$\!\!\!$$-$42.7& Fe II  (74)&$\!\!\!$ 2.60 &           \\
 4798.54 & 0.28 &$\!\!\!$ 1.22 &$\!\!\!$$-$42.5& Ti II  (17)&$\!\!\!$  1.32 &                  & 6516.05 &  2.44 &$\!\!\!$ 2.00 &$\!\!\!$$-$41.0& Fe II  (74)&$\!\!\!$ 2.48 &           \\
 4805.11 & 0.58 &$\!\!\!$ 1.30 &$\!\!\!$$-$41.2& Ti II  (92)&$\!\!\!$  1.75 &                  & 6562.82 &105.00 &$\!\!\!$15.50 &$\!\!\!$ $-$9.1& H   I  ( 1)&$\!\!\!$ 5.91 &           \\
 4824.13 & 1.07 &$\!\!\!$ 1.51 &$\!\!\!$$-$45.4& Cr II  (30)&$\!\!\!$  2.19 &  b:              & 6604.60 &  0.30 &$\!\!\!$ 1.11 &$\!\!\!$$-$31.3& Sc II  (19)&$\!\!\!$ 2.76 &           \\
 4836.22 & 0.42 &$\!\!\!$ 1.23 &$\!\!\!$$-$43.4& Cr II  (30)&$\!\!\!$  2.02 &                  & 6613.63 &  0.17 &$\!\!\!$ 0.84 &$\!\!\!$$-$26.3& DIB        &$\!\!\!$ 0.97 &           \\
 4848.24 & 0.43 &$\!\!\!$ 1.23 &$\!\!\!$$-$66.2& Cr II  (30)&$\!\!\!$  1.73 &  b w/H$\beta$    & 6660.64 &  0.14 &$\!\!\!$ 0.93 &$\!\!\!$$-$27.5& DIB        &$\!\!\!$ 2.69 &           \\
 4861.33 & 6.58 &$\!\!\!$ 0.25 &$\!\!\!$$-$423.3& H  I  ( 1)&$\!\!\!$  8.75 &                  & 6699.32 &  0.02 &$\!\!\!$ 0.98 &$\!\!\!$$-$34.9& DIB        &$\!\!\!$ 0.87 &           \\
 4861.33 &16.30 &$\!\!\!$ 4.75 &$\!\!\!$$-$11.1& H   I  ( 1)&$\!\!\!$  3.50 &                  & 6717.00 &  0.00 &$\!\!\!$ 1.13 &$\!\!\!$$-$451.1&S  II  ( 2)&$\!\!\!$ 0.00 &           \\
\noalign{\smallskip}\hline
\smallskip
\end{tabular}
} \addtocounter{table}{-1}
\end{table*}

\begin{table*}
\caption{Line identification in the optical spectrum of V645\,Cyg.
Continued}
{\tiny
\begin{tabular}{crrrlclcrrrlcl}
\hline\noalign{\smallskip}
$\lambda$& EW    & I/I$_c$& V$_r$& ID      & FWHM  & Comment &$\lambda$& EW    & I/I$_c$& V$_r$& ID      & FWHM  & Comment \\
\noalign{\smallskip}\hline\noalign{\smallskip}
 4876.41 & 0.83 &$\!\!\!$ 1.35 &$\!\!\!$$-$50.5& Cr II  (30)&$\!\!\!$  2.35 &                  & 6717.00 &  0.00 &$\!\!\!$ 1.13 &$\!\!\!$$-$115.7&S  II  ( 2)&$\!\!\!$ 0.98 &           \\
 4884.57 & 0.52 &$\!\!\!$ 1.24 &$\!\!\!$$-$80.5& Cr II  (30)&$\!\!\!$  1.92 &                  & 6717.69 &  0.10 &$\!\!\!$ 1.06 &$\!\!\!$$-$21.9& Ca  I  ( 3)&$\!\!\!$ 2.05 &  +p       \\
 4899.90 & 0.33 &$\!\!\!$ 1.22 &$\!\!\!$$-$22.7& Fe II  (30)&$\!\!\!$  1.48 &  P Cyg:, ID:     & 6731.30 &  0.00 &$\!\!\!$ 1.17 &$\!\!\!$$-$400.7&S  II  ( 2)&$\!\!\!$ 0.00 &           \\
 4911.20 & 0.54 &$\!\!\!$ 1.27 &$\!\!\!$$-$45.2& Ti II (114)&$\!\!\!$  1.89 &  b               & 6731.30 &  0.00 &$\!\!\!$ 1.29 &$\!\!\!$$-$97.1& S  II  ( 2)&$\!\!\!$ 1.43 &           \\
 4923.92 & 2.60 &$\!\!\!$ 0.57 &$\!\!\!$$-$396.6& Fe II (42)&$\!\!\!$  5.67 &                  & 6828.50 &  0.15 &$\!\!\!$ 1.06 &$\!\!\!$$-$51.4& C   I  (21)&$\!\!\!$ 2.64 &  ID:      \\
 4923.92 & 3.44 &$\!\!\!$ 2.32 &$\!\!\!$$-$40.8& Fe II  (42)&$\!\!\!$  2.50 &                  & 7004.60 &  0.16 &$\!\!\!$ 1.07 &$\!\!\!$$-$44.5& Ti  I (256)&$\!\!\!$ 1.81 &  ID:      \\
 4934.09 & 0.35 &$\!\!\!$ 1.17 &$\!\!\!$$-$53.5& Ba II  ( 1)&$\!\!\!$  2.21 &                  & 7155.14 &  0.00 &$\!\!\!$ 1.15 &$\!\!\!$$-$86.4& Fe II  (14)&$\!\!\!$ 0.00 &           \\
 4957.15 & 0.16 &$\!\!\!$ 1.19 &$\!\!\!$$-$18.8& Ba II  (10)&$\!\!\!$  0.95 &  ID:             & 7214.69 &  0.34 &$\!\!\!$ 1.12 &$\!\!\!$$-$37.8& Fe II  (30)&$\!\!\!$ 2.72 &  ID:      \\
 4993.35 & 0.83 &$\!\!\!$ 1.35 &$\!\!\!$$-$42.7& Fe II  (36)&$\!\!\!$  2.92 &                  & 7222.39 &  0.00 &$\!\!\!$ 1.20 &$\!\!\!$$-$28.7& Fe II  (73)&$\!\!\!$ 0.00 &  c        \\
 5018.43 & 3.36 &$\!\!\!$ 0.59 &$\!\!\!$$-$417.9& Fe II  (42)&$\!\!\!$  7.47 &                 & 7291.46 &  0.00 &$\!\!\!$ 1.52 &$\!\!\!$$-$37.4& Ca II  ( 1)&$\!\!\!$ 0.00 &  c        \\
 5018.43 & 3.51 &$\!\!\!$ 2.32 &$\!\!\!$$-$37.7& Fe II  (42)&$\!\!\!$  2.58 &                  & 7307.97 &  0.00 &$\!\!\!$ 1.23 &$\!\!\!$$-$27.5& Fe II  (73)&$\!\!\!$ 0.00 &  c        \\
 5031.02 & 0.25 &$\!\!\!$ 1.17 &$\!\!\!$$-$42.3& Sc II  (23)&$\!\!\!$  1.31 &                  & 7320.70 &  0.00 &$\!\!\!$ 1.20 &$\!\!\!$$-$41.0& Fe II  (73)&$\!\!\!$ 0.00 &  c        \\
 5129.14 & 0.00 &$\!\!\!$ 1.47 &$\!\!\!$$-$45.0& Ti II  (86)&$\!\!\!$  0.00 &  b               & 7323.88 &  0.00 &$\!\!\!$ 1.41 &$\!\!\!$$-$42.6& Ca II  ( 1)&$\!\!\!$ 0.00 &  +p       \\
 5146.06 & 0.16 &$\!\!\!$ 1.12 &$\!\!\!$$-$39.7& O   I  (28)&$\!\!\!$  1.15 &                  & 7376.46 &  0.00 &$\!\!\!$ 1.17 &$\!\!\!$$-$61.4& Fe II  (  )&$\!\!\!$ 0.00 &  +p       \\
 5154.06 & 1.44 &$\!\!\!$ 1.52 &$\!\!\!$$-$47.2& Ti II  (70)&$\!\!\!$  3.22 &                  & 7462.38 &  1.14 &$\!\!\!$ 1.40 &$\!\!\!$$-$42.2& Fe II  (73)&$\!\!\!$ 2.83 &           \\
 5169.03 & 3.30 &$\!\!\!$ 0.47 &$\!\!\!$$-$363.9& Fe II  (42)&$\!\!\!$  5.86 &                 & 7533.42 &  0.34 &$\!\!\!$ 1.10 &$\!\!\!$$-$53.8& Fe II  (72)&$\!\!\!$ 3.06 &           \\
 5169.03 & 0.00 &$\!\!\!$ 2.28 &$\!\!\!$$-$38.3& Fe II  (42)&$\!\!\!$  0.00 &                  & 7698.98 &  0.00 &$\!\!\!$ 0.78 &$\!\!\!$$-$42.9& K   I  ( 1)&$\!\!\!$ 0.00 &  +2f      \\
 5172.68 & 0.00 &$\!\!\!$ 1.83 &$\!\!\!$$-$60.9& Mg  I  ( 2)&$\!\!\!$  0.00 &  +p              & 7698.98 &  0.00 &$\!\!\!$ 0.56 &$\!\!\!$$-$31.6& K   I  ( 1)&$\!\!\!$ 0.00 &           \\
 5183.60 & 0.00 &$\!\!\!$ 1.59 &$\!\!\!$$-$40.5& Mg  I  ( 2)&$\!\!\!$  0.00 &                  & 7698.98 &  0.00 &$\!\!\!$ 0.92 &$\!\!\!$$-$14.0& K   I  ( 1)&$\!\!\!$ 0.00 &           \\
 5188.70 & 0.00 &$\!\!\!$ 1.47 &$\!\!\!$$-$45.1& Ti II  (70)&$\!\!\!$  0.00 &                  & 7711.71 &  1.75 &$\!\!\!$ 1.52 &$\!\!\!$$-$43.6& Fe II  (73)&$\!\!\!$ 3.27 &           \\
 5197.57 & 2.53 &$\!\!\!$ 2.13 &$\!\!\!$$-$40.4& Fe II  (49)&$\!\!\!$  2.38 &                  & 7774.00 &  5.15 &$\!\!\!$ 0.59 &$\!\!\!$$-$323.0& O   I  ( 1)&$\!\!\!$12.70 &          \\
 5226.53 & 1.22 &$\!\!\!$ 1.59 &$\!\!\!$$-$34.4& Ti II  (70)&$\!\!\!$  2.00 &                  & 8327.05 &  0.15 &$\!\!\!$ 1.18 &$\!\!\!$$-$47.5& Fe  I  (60)&$\!\!\!$ 1.67 &           \\
 5234.62 & 0.00 &$\!\!\!$ 2.05 &$\!\!\!$$-$44.7& Fe II  (49)&$\!\!\!$  1.97 &  b:              & 8387.78 &  0.00 &$\!\!\!$ 1.27 &$\!\!\!$$-$37.5& Fe  I  (60)&$\!\!\!$ 0.00 &  b w/Pa20 \\
 5254.92 & 0.62 &$\!\!\!$ 1.30 &$\!\!\!$$-$37.7& Fe II  (49)&$\!\!\!$  1.92 &  b               & 8413.32 &  1.27 &$\!\!\!$ 1.24 &$\!\!\!$$-$51.0& H   I  ( 1)&$\!\!\!$ 5.89 &  b: Pa19  \\
 5261.71 & 0.00 &$\!\!\!$ 1.14 &$\!\!\!$$-$29.1& Ca  I  (22)&$\!\!\!$  1.64 &  +f              & 8446.35 &  0.00 &$\!\!\!$ 1.96 &$\!\!\!$$-$55.7& O   I  ( 4)&$\!\!\!$ 3.54 &  b:       \\
 5264.80 & 0.50 &$\!\!\!$ 1.25 &$\!\!\!$$-$45.6& Fe II  (48)&$\!\!\!$  1.88 &                  & 8467.26 &  1.27 &$\!\!\!$ 1.26 &$\!\!\!$$-$39.3& H   I  ( 1)&$\!\!\!$ 4.89 &           \\
 5275.99 & 0.31 &$\!\!\!$ 1.19 &$\!\!\!$$-$408.9& Fe II  (49)&$\!\!\!$  1.55 &  bp:            & 8498.02 & 15.92 &$\!\!\!$ 4.64 &$\!\!\!$$-$35.6& Ca II  ( 2)&$\!\!\!$ 4.00 &  +P16 PCyg\\
 5275.99 & 1.94 &$\!\!\!$ 1.83 &$\!\!\!$$-$48.3& Fe II  (49)&$\!\!\!$  2.32 &                  & 8542.09 &  2.82 &$\!\!\!$ 0.78 &$\!\!\!$$-$445.0& Ca II ( 2)&$\!\!\!$11.77 &  +P15 PCyg\\
 5284.09 & 1.33 &$\!\!\!$ 1.68 &$\!\!\!$$-$44.9& Fe II  (41)&$\!\!\!$  1.89 &                  & 8542.09 & 18.40 &$\!\!\!$ 4.56 &$\!\!\!$$-$19.0& Ca II  ( 2)&$\!\!\!$ 5.08 &           \\
 5305.85 & 0.07 &$\!\!\!$ 1.06 &$\!\!\!$$-$40.1& Cr II  (24)&$\!\!\!$  0.90 &                  & 8662.14 &  2.16 &$\!\!\!$ 0.85 &$\!\!\!$$-$417.7& Ca II ( 2)&$\!\!\!$13.39 &           \\
 5316.78 & 0.99 &$\!\!\!$ 0.77 &$\!\!\!$$-$378.0& Fe II  (48)&$\!\!\!$  4.24 &                 & 8662.14 & 16.30 &$\!\!\!$ 4.20 &$\!\!\!$$-$26.0& Ca II  ( 2)&$\!\!\!$ 4.93 &  +P13 PCyg\\
 5316.78 & 2.80 &$\!\!\!$ 2.29 &$\!\!\!$$-$53.0& Fe II  (48)&$\!\!\!$  2.43 &                  & 8688.63 &  0.48 &$\!\!\!$ 1.20 &$\!\!\!$$-$43.5& Fe  I  (60)&$\!\!\!$ 2.38 &           \\
 5362.86 & 1.77 &$\!\!\!$ 1.83 &$\!\!\!$$-$45.3& Fe II  (48)&$\!\!\!$  2.12 &  n, PCyg         & 8727.40 &  0.27 &$\!\!\!$ 1.12 &$\!\!\!$$-$69.5& C   I  ( 3)&$\!\!\!$ 2.22 &  ID:      \\
 5526.80 & 0.74 &$\!\!\!$ 1.33 &$\!\!\!$$-$42.3& Sc II  (31)&$\!\!\!$  2.33 &                  & 8750.00 &  2.00 &$\!\!\!$ 1.44 &$\!\!\!$$-$34.6& Pa 12      &$\!\!\!$ 5.02 &  PCyg     \\
 5534.86 & 1.27 &$\!\!\!$ 1.61 &$\!\!\!$$-$46.1& Fe II  (55)&$\!\!\!$  2.10 &                  & 8806.77 &  0.97 &$\!\!\!$ 1.22 &$\!\!\!$$-$44.3& Mg  I  ( 7)&$\!\!\!$ 4.49 &  P Cyg:   \\
 5577.34 & 0.79 &$\!\!\!$ 7.58 &$\!\!\!$  10.8& O    I  ( 3)&$\!\!\!$  0.09 &  t               & 8824.22 &  0.55 &$\!\!\!$ 1.16 &$\!\!\!$$-$39.8& Fe  I  (60)&$\!\!\!$ 3.36 &           \\
 5586.76 & 0.23 &$\!\!\!$ 1.10 &$\!\!\!$$-$33.3& Fe  I (686)&$\!\!\!$  2.35 &                  & 8862.79 &  1.56 &$\!\!\!$ 1.43 &$\!\!\!$$-$50.4& Pa 11      &$\!\!\!$ 3.49 &      P Cyg\\
 5615.65 & 0.17 &$\!\!\!$ 1.10 &$\!\!\!$$-$47.6& Fe  I (686)&$\!\!\!$  1.84 &                  & 8927.36 &  1.11 &$\!\!\!$ 1.14 &$\!\!\!$$-$59.8& Ca II  (  )&$\!\!\!$ 4.03 &  c        \\
\noalign{\smallskip}\hline
\smallskip
\end{tabular}
}
\end{table*}
\newpage
\begin{list}{}
\item Column information: 1 -- laboratory wavelength of the line, 2
-- equivalent width in \AA, 3 -- peak intensity in continuum units,
4 -- heliocentric radial velocity of the line peak from fitting to a
Gaussian, 5 -- line identification with the multiplet number
according to Coluzzi (\cite{col93}), 6 -- full-width at half maximum
in \AA, 7 -- comment.
\item Comment notation: b -- blend, b: -- possible blend, c -- contaminated with telluric
spectrum, n -- noisy, t -- telluric line; ID: -- uncertain
identification, +f -- blend with the following line, +p -- blend
with the previous line, P\,Cyg -- the line has a P\,Cyg type
profile; P\,Cyg: -- weak or noisy P\,Cyg type absorption.
\item Comments to Table content: zeros in columns mean that the
parameter cannot be measured due to noisiness or blending.
Parameters of the P\,Cyg type absorption components are measured
separately from the emission components of the same line only in
cases when the absorptions are clearly seen. Equivalent width and
FWHM for double-peaked profiles (e.g., [S {\sc II}] $\lambda$6717
\AA\ and $\lambda$6731 \AA).
\end{list}

\end{document}